\documentclass[12pt]{article}

\usepackage{bm}
\usepackage{empheq}
\usepackage{epsfig}
\usepackage{graphicx}
\usepackage{epstopdf}
\usepackage{mathtools}
\usepackage{amssymb}
\usepackage{amsmath}

\makeindex
\def\disp{\displaystyle}

\def\be{\begin{equation}}
\def\ee{\end{equation}}
\def\bea{\begin{eqnarray}}
\def\eea{\end{eqnarray}}
\def\beaN{\begin{eqnarray*}}
\def\eeaN{\end{eqnarray*}}
\def\ed{\end{document}}
\def\bit{\begin{itemize}}
\def\eit{\end{itemize}}

\def\sig{\sigma}
\def\Sig{\Sigma}
\def\lam{\lambda}
\def\Lam{\Lambda}
\def\Del{\Delta}
\def\del{\delta}

\def\k{\kappa}

\def\alf{\alpha}

\def\3nabla{{\stackrel{\scriptscriptstyle 3}{\nabla}}}
\def\2nabla{{\stackrel{\scriptscriptstyle 2}{\nabla}}}

\def\di{\partial}

\def\Lix{\pounds_\xi}

\def\half{{\textstyle{1 \over 2}}}

\def\~{\tilde}
\def\lag{{{\cal L}}}

\def\m{\label}
\def\l{\left}
\def\r{\right}

\def\goto{\rightarrow}
\def\Bar{\overline}

\def\citep{\cite}

\newcommand{\gog}{\mathfrak{g}}
\newcommand{\goh}{\mathfrak{h}}

\usepackage{cite}

\newcommand{\nc}{\newcommand}

\nc{\bse}{\begin{equation*}}
\nc{\ese}{\end{equation*}}
\nc{\ba}{\begin{array}}
\nc{\ea}{\end{array}}
\nc{\bal}{\begin{align}}
\nc{\eal}{\end{align}}
\nc{\bi}{\begin{description}}
\nc{\ei}{\end{description}}
\nc{\Def}{=}
\nc{\sign}{\mathrm{sign}\;}
\nc{\diagR}{\mathrm{diag}\;}
\nc{\constR}{\mathrm{const}}
\nc{\Tr}{\mathrm{Tr}\,}
\nc{\id}{\mathrm{id}}
\nc{\eq}{\equiv}
\nc{\we}{\wedge}
\nc{\ra}{\rightarrow}
\nc{\bfrac}{\disp\frac}

\nc{\bfpa}{{\bm{\partial}}}                       
\nc{\pa}[1]{{\partial_{#1}}{}}                    
\nc{\pau}[1]{{\partial^{#1}}{}}                   
\nc{\alp}{\alpha}
\nc{\bet}{\beta}
\nc{\gam}{\gamma}
\nc{\eps}{\epsilon}
\nc{\veps}{\varepsilon}
\nc{\zet}{\zeta}
\nc{\tet}{\theta}
\nc{\vtet}{\vartheta}
\nc{\iot}{\iota}
\nc{\kap}{\kappa}
\nc{\vkap}{\varkappa}
\nc{\vpi}{\varpi}
\nc{\vrho}{\varrho}
\nc{\vsig}{\varsigma}
\nc{\ups}{\upsilon}
\nc{\vphi}{\varphi}
\nc{\ome}{\omega}

\nc{\Gam}{\Gamma}
\nc{\Tet}{\Theta}
\nc{\Ups}{\Upsilon}
\nc{\Ome}{\Omega}

\nc{\BL}[1]{\bm{#1}}

\nc{\bfa}{\BL{a}}
\nc{\bfb}{\BL{b}}
\nc{\bfc}{\BL{c}}
\nc{\bfd}{\BL{d}}
\nc{\bfe}{\BL{e}}
\nc{\bff}{\BL{f}}
\nc{\bfg}{\BL{g}}
\nc{\bfh}{\BL{h}}
\nc{\bfi}{\BL{i}}
\nc{\bfj}{\BL{j}}
\nc{\bfk}{\BL{k}}
\nc{\bfl}{\BL{l}}
\nc{\bfm}{\BL{m}}
\nc{\bfn}{\BL{n}}
\nc{\bfo}{\BL{o}}
\nc{\bfp}{\BL{p}}
\nc{\bfq}{\BL{q}}
\nc{\bfr}{\BL{r}}
\nc{\bfs}{\BL{s}}
\nc{\bft}{\BL{t}}
\nc{\bfu}{\BL{u}}
\nc{\bfv}{\BL{v}}
\nc{\bfw}{\BL{w}}
\nc{\bfx}{\BL{x}}
\nc{\bfy}{\BL{y}}
\nc{\bfz}{\BL{z}}

\nc{\bfA}{\BL{A}}
\nc{\bfB}{\BL{B}}
\nc{\bfC}{\BL{C}}
\nc{\bfD}{\BL{D}}
\nc{\bfE}{\BL{E}}
\nc{\bfF}{\BL{F}}
\nc{\bfG}{\BL{G}}
\nc{\bfH}{\BL{H}}
\nc{\bfI}{\BL{I}}
\nc{\bfJ}{\BL{J}}
\nc{\bfK}{\BL{K}}
\nc{\bfL}{\BL{L}}
\nc{\bfM}{\BL{M}}
\nc{\bfN}{\BL{N}}
\nc{\bfO}{\BL{O}}
\nc{\bfP}{\BL{P}}
\nc{\bfQ}{\BL{Q}}
\nc{\bfR}{\BL{R}}
\nc{\bfS}{\BL{S}}
\nc{\bfT}{\BL{T}}
\nc{\bfU}{\BL{U}}
\nc{\bfV}{\BL{V}}
\nc{\bfW}{\BL{W}}
\nc{\bfX}{\BL{X}}
\nc{\bfY}{\BL{Y}}
\nc{\bfZ}{\BL{Z}}

\nc{\bfalp}{\bm{\alp}}
\nc{\bfbet}{\bm{\bet}}
\nc{\bfgam}{\bm{\gam}}
\nc{\bfdel}{\bm{\del}}
\nc{\bfeps}{\bm{\eps}}
\nc{\bfveps}{\bm{\veps}}
\nc{\bfzet}{{\bm{\zet}}}
\nc{\bfeta}{\bm{\eta}}
\nc{\bftet}{\bm{\tet}}
\nc{\bfvtet}{\bm{\vtet}}
\nc{\bfiot}{\bm{\iot}}
\nc{\bfkap}{\bm{\kap}}
\nc{\bfvkap}{\bm{\vkap}}
\nc{\bflam}{\bm{\lam}}
\nc{\bfmu}{\bm{\mu}}
\nc{\bfnu}{\bm{\nu}}
\nc{\bfxi}{\bm{\xi}}
\nc{\bfpi}{\bm{\pi}}
\nc{\bfvpi}{\bm{\vpi}}
\nc{\bfrho}{\bm{\rho}}
\nc{\bfvrho}{\bm{\vrho}}
\nc{\bfsig}{\bm{\sig}}
\nc{\bfvsig}{\bm{\vsig}}
\nc{\bftau}{\bm{\tau}}
\nc{\bfups}{\bm{\ups}}
\nc{\bfphi}{\bm{\phi}}
\nc{\bfvphi}{\bm{\vphi}}
\nc{\bfchi}{\bm{\chi}}
\nc{\bfpsi}{\bm{\psi}}
\nc{\bfome}{\bm{\ome}}

\nc{\bfGam}{\bm{\Gam}}
\nc{\bfDel}{\bm{\Del}}
\nc{\bfTet}{\bm{\Tet}}
\nc{\bfLam}{\bm{\Lam}}
\nc{\bfXi}{\bm{\Xi}}
\nc{\bfPi}{\bm{\Pi}}
\nc{\bfSig}{\bm{\Sig}}
\nc{\bfUps}{\bm{\Ups}}
\nc{\bfPhi}{\bm{\Phi}}
\nc{\bfPsi}{\bm{\Psi}}
\nc{\bfOme}{\bm{\Ome}}

\begin{document}

\centerline{\bf Field-theoretical construction of currents and superpotentials}
\centerline{\bf in Lovelock gravity}

\smallskip

\centerline{\it A.N.Petrov}

\smallskip

\centerline{\it Moscow MV Lomonosov State university, Sternberg
Astronomical institute,} \centerline{\it
 Universitetskii pr., 13, Moscow, 119992,
RUSSIA }

\smallskip

\centerline{ e-mail: alex.petrov55@gmail.com}

\smallskip

telephone number: +7(495)7315222

\smallskip

fax: +7(495)9328841

\smallskip

PACS numbers: 04.50.+h, 04.20.Cv, 04.20.Fy

\begin{abstract}
Conserved currents and related superpotentials for perturbations on arbitrary backgrounds in the Lovelock theory are constructed. We use the Lagrangian based field-theoretical method where perturbations are considered as dynamical fields propagating on a given background. Such a formulation is exact (not approximate) and equivalent to the theory in the original metric form. From the very start, using Noether theorem, we derive the Noether-Klein identities and adopt them for the purposes of the current work. Applying these identities in the framework of Lovelock theory, we construct conserved currents, energy-momentum tensors out of them, and related superpotentials with arbitrary displacement vectors, not restricting to Killing vectors. A comparison with the well known Abbott-Deser-Tekin approach is given. The developed general formalism is applied to give conserved quantities for perturbations on anti-de Sitter (AdS) backgrounds. As a test we calculate mass of the
Schwarzschild-AdS black hole in the Lovelock theory in arbitrary $D$ dimensions. Proposals for future applications are presented.
\end{abstract}

\eject

\section{Introduction}
\m{Introduction}

The popularity of the Lovelock theory \cite{Lovelock} has steadily risen over the last two-three decades. Most of the attention has been paid to constructing black hole solutions of different kinds and to studying stability and thermodynamic properties of such solutions, see, for example, \cite{Troncoso+_2000,Cai_Ohta_2006,Taves_2014,Aros_Estrada_2019} and numerous references therein. For a study of cosmological solutions in the framework of the Lovelock theory see, for example, \cite{Pavluchenko_Toporensky_2014}. The main goal of the present paper is to give a formalism for constructing conserved quantities in the Lovelock theory that can be an effective tool for analyzing such solutions, sometimes exotic ones.

In the current work, we develop a construction of conserved
quantities for perturbations on a fixed background in the framework of the programm started
in \cite{Petrov2009,Petrov2009a,Petrov2011} for the Einstein-Gauss-Bonnet (EGB) gravity.
The program presents three approaches simultaneously. They are
1) {\em canonical} method based on a direct application of the
Noether theorem, for
4-dimensional general relativity (4D GR) see \cite{KBL}; 2) {\em
Belinfante symmetrization} method, for 4D GR see \cite{PK}; 3) {\em field-theoretical} method that we apply here.

The field-theoretical approach begins from the linearization in metric perturbations  of the
gravitational equations. All the other terms are replaced to the right hand side and treated as an
effective energy-momentum, which is a source for the linear part
(see, for example, the book \cite{Weinberg-book}). This construction has been represented in the form of the exact (without approximation) {\em Lagrangian based field theory} for
metrical perturbations  by Deser on flat backgrounds \cite{[11]} and anti-de Sitter (AdS) backgrounds  \cite{Deser87}.
His model was further developed for arbitrary curved backgrounds (including non-vacuum ones) in our works \cite{GPP,GP87,[15]} with constructing currents and superpotentials including arbitrary displacement vectors (not only Killing ones) \cite{CQG_DT,Petrov2008}, and many applications have been provided; for a review see chapters 2 - 4
in the book \cite{Petrov+_2017}. In the present paper, we use namely the {\em Lagrangian based} method, the main advantage of which is the possibility to apply the Noether theorem in the standard way.

All the three approaches are based on the well known
form of conservation laws, where a conserved current is expressed
through a divergence of a related superpotential. In literature, as a rule, the main attention is paid to constructing superpotentials. They give a possibility to describe global characteristics of physical systems such as conserved charges, quasi-local quantities \cite{Szabados_2009}, {\em etc}. Following this necessity, in \cite{Petrov2009}, we have suggested superpotentials of all
the three aforementioned kinds in the EGB gravity in the most general form. They are constructed  for arbitrary perturbations on arbitrary curved backgrounds, with arbitrary displacement vectors, in the exact form.

Currents, unlike superpotentials, describe local properties of physical systems. However, to the best of our knowledge, local characteristics usually are described by energy-momentum tensors (or pseudotensors) only. As a rule, it is not enough, at least, if perturbations are propagated in a curved non-vacuum background spacetime. Thus, one needs  to construct conserved covariant currents for perturbations. A structure of such currents has to permit to take into account background effects, to calculate angular momenta, {\em etc}. With regards to the EGB gravity, already, in \cite{Petrov2011}, we have constructed the currents of the two types, canonical and Belinfante corrected ones, corresponding to the superpotentials of the related types in  \cite{Petrov2009}. These new conserved quantities in EGB gravity have been tested and applied \cite{Petrov2009,Petrov2009a,Petrov2011} as well.

To continue the program it is necessary to construct currents of the third type that is in the framework of the field-theoretical approach for perturbations in the EGB gravity. In the book \cite{Petrov+_2017} we have suggested general principles for constructing such currents. Here, we develop them in detail. Moreover, in the present paper, we construct conserved currents and superpotentials in the most general form in the Lovelock theory of an arbitrary order \cite{Lovelock}. The EGB gravity is a particular case. To test the new expressions for conserved quantities we calculate the mass of the Schwarzschild-AdS black hole in the Lovelock theory in arbitrary $D$ dimensions.

The paper is organized as follows. In section \ref{PerturbedDTheory}, we present the main properties of the Lagrangian based field-theoretical reformulation of an arbitrary metric theory. In section \ref{NIdentities}, making the use of the Noether theorem, we derive the Noether-Klein identities \cite{Petrov_Lompay_2013} and adopt them for a direct application in the field-theoretical formulation of the Lovelock theory. In section \ref{ConservedCurrents}, the results of section \ref{NIdentities} are used to construct conserved currents, energy-momentum tensors and related superpotentials  in the framework of the Lovelock theory in a general form without structured expressions. In section \ref{Lovelock}, we concretize expressions for currents and superpotentials in the Lovelock theory for perturbations on arbitrary curved, both vacuum and non-vacuum,  backgrounds. In section \ref{S-AdS-mass}, we present currents and superpotentials for the Lovelock theory for perturbations on AdS backgrounds and calculate mass of the Schwarzschild-AdS black hole. In section 7, we discuss the obtained results, stress and accent their advantages with respect to other approaches and announce future applications.

Below, in the remainder of this section, we introduce primary notations:

\noindent - $\psi^A,~{P}_B, \ldots$ - sets of tensor densities of arbitrary ranks and weights
with the collective indices $A,~ B,\ldots$ in a piggyback notation;

\noindent - $\bar {\psi^A}$ - the ``bar'' above $\psi^A$ means a background value of $\psi^A$;

\noindent - ${\bm t}_\sig{}^\mu, {\bm m}_\sig{}^{\mu\nu}, \ldots$ - notations in calligraphic boldface for small letters, if they represent quantities of mathematic weight +1. For example, ${\bm t}_\sig{}^\mu$ could be a density  expressed with the use of the tensor ${t}_\sig{}^\mu$:  ${\bm t}_\sig{}^\mu = \sqrt{-\bar g}{t}_\sig{}^\mu$, or ${\bm t}_\sig{}^\mu$  could be a density itself, {\em etc};

\noindent - ${\cal L},~ {\cal U}_\sig{}^{\mu}, \ldots$ - the capital calligraphic letters
denote geometric quantities of weight +1 analogously to previous item;

\noindent - ${\xi^\alf}$ and ${\bar\xi^\alf}$ - arbitrary displacement vectors and Killing vectors, respectively, in a spacetime;

\noindent -  $g_{\mu\nu}$ and $\bar g_{\mu\nu}$ - the dynamical and background metrics;

\noindent - $g ={\det}\,g_{\mu\nu}$ and $\bar g =\det\,\bar g_{\mu\nu}$ - the determinants of dynamical and background metrics;

\noindent - the indices of tensor fields on the physical manifolds or on the background manifolds are lowered and raised with the use of $g_{\alf\beta}$ or $\bar g_{\alf\beta}$ and their inverses, respectively;

\noindent - $R^\rho{}_{\alf\sig\beta}$ and $\bar R^\rho{}_{\alf\sig\beta}$, $R_{\alf\beta}$ and $\bar R_{\alf\beta}$, and $R$ and $\bar R$ - the Riemannian, Ricci tensors and the Ricci scalars in physical and background spacetimes, respectively;

\noindent - ${\di{\psi^A} }/{\di x^\alf}= \di_\alf \psi^A = {\psi}^{A}{}_{,\alf}$ -  the partial derivative;

\noindent - $\nabla_\alf {\psi}^A$ and $\bar\nabla_\alf {\psi}^A$ -  the covariant derivatives of ${\psi}^A$ compatible with $g_{\mu\nu}$ and $\bar g_{\mu\nu}$, respectively;

\noindent - the Lagrange derivative of the quantity ${\psi}^A={\psi}^A(q^B;q^B{}_{,\alf};q^B{}_{,\alf\beta})$ is
 $$\displaystyle\frac{\del{\psi}^A}{\del q^B} = \frac{\di{\psi}^A}{\di q^B} - \di_\alf\l(\frac{\di{\psi}^A}{\di q^B{}_{,\alf}}\r) + \di_\alf\di_\beta\l(\frac{\di{\psi}^A}{\di q^B{}_{,\alf\beta}}\r)\,;$$

\noindent - $\l.{\psi}^A\r|^\alf_\beta$ is a permutation linear operator depending on the transformation properties of ${\psi}^A$, for example, for the tensor density $\psi^A = {\bm t}_\sig{}^\mu$ one has $\l. {\bm t}_\sig{}^\mu \r|^\gamma_\beta = - {\bm t}_\sig{}^\mu \delta^\gamma_\beta + {\bm t}_\sig{}^\gamma \delta^\mu_\beta  - {\bm t}_\beta{}^\mu \delta^\gamma_\sig$;

\noindent - $\Lix {\psi}^A = -\xi^\alf \psi^A{}_{,\alf} + \xi^\beta{}_{,\alf}\l.{\psi}^A\r|^\alf_\beta$ - the Lie derivative of the quantity ${\psi}^A$ along $\xi^\alf$.

\section{The Lagrangian based field-theoretical approach in metric theories}
 \m{PerturbedDTheory}
 \setcounter{equation}{0}

In the present section, following \cite{GPP,[15]}, we present the main properties of the Lagrangian based field-theoretical method in metric theories, which are used in the current paper.
 Consider a $D$-dimensional
metric theory with the Lagrangian:
 \be
 \lag(g^a,\Phi^A) = - \frac{1}{2\k}\lag_{g}(g^a) + \lag_{m}(g^a,\Phi^A)\,.
 \m{lag-g}
 \ee
For the sake of generality, the pure metric part $\lag_{g}$ depends on
a metric variable $g^a$ that is thought as a one from the set
  \be
 g^{a} = \l\{g^{\mu \nu },~g_{\mu
\nu },~\sqrt{-g}g^{\mu \nu },~\sqrt{-g}g_{\mu \nu },~(-g)g^{\mu \nu
},~\ldots\r\}\,. \m{(1)}
 \ee
The matter part $\lag_{m}$ in (\ref{lag-g}) depends on $\Phi^A$ --- generalized matter variables interacting with $g^a$. The variation of the Lagrangian
(\ref{lag-g}) with respect to $g^a$ leads to the field equations:
  \be
\frac{\delta \lag_{g}}{\delta g^a} = 2\k \frac{\delta
\lag_{m}}{\delta g^a}\qquad \goto \qquad \frac{\delta
\lag_{g}}{\delta g^{\mu\nu}} = \k {\cal T}_{\mu\nu}\, . \m{ddd}
 \ee
The last equation is the gravitational equations in the standard form obtained after contracting the first one with $\di g^a/\di g^{\mu\nu}$.
 Variation of (\ref{lag-g}) with
respect to $\Phi^A$ gives corresponding matter equations. The
background gravitational equations are in the form
 \be
 \frac{\delta
\bar\lag_{g}}{\delta \bar g^a} = 2\k \frac{\delta \bar
\lag_{m}}{\delta \bar g^a}\qquad \goto \qquad \frac{\delta {\bar
\lag}_{g}}{\delta \bar g^{\mu\nu}} = \k \bar {\cal T}_{\mu\nu}\,,
 \m{bbb}
 \ee
where the background Lagrangian is defined by the barred procedure
in (\ref{lag-g}): $\bar \lag = \lag(\bar g^a,\bar\Phi^A)$.
The background matter equations are derived analogously.  It is assumed that background fields $\bar g^a$ and $\bar\Phi^A$ are known (fixed) and satisfy the
background equations.

A physical system described by equations (\ref{ddd}) can be considered as a perturbed one
with respect to a background system with the equations (\ref{bbb}). Thus, from the start we
decompose metric and matter variables in (\ref{lag-g}) into the
background (barred) parts and the dynamic variables (perturbations)
$h^a$ and $\phi^A$:
 \bea
 g^a &=& \bar g^a + h^a\, ,
 \m{g-Dec}\\
 \Phi^A &=& \bar \Phi^A + \phi^A\, .
  \m{Phi-Dec}
 \eea
At the end of this section we explain
why it is important to examine the different definitions $h^a$ related to different variables in (\ref{(1)}).
After that we represent the perturbed system by the Lagrangian:
 \be
\lag^{dyn}(\bar g,\bar\Phi;\,h,\phi) = \lag (\bar g+h,\,\bar
\Phi+\phi ) - h^a \frac{\delta \bar \lag}{\delta \bar g^a} - \phi^A
\frac{\delta \bar \lag}{\delta \bar \Phi^A}- \bar \lag\,.
 \m{lag}
 \ee
We call it the
{\em dynamic Lagrangian} because perturbations $h^a$ and $\phi^A$ in (\ref{lag}) are treated as the {\em dynamic variables}. Note that we do not use
the original Lagrangian, $\lag(g,\Phi) = \lag(\bar
g+h,\,\bar \Phi+\phi)$, after substituting the decompositions (\ref{g-Dec}) and (\ref{Phi-Dec}). Thus, the field-theoretical method applied in the current paper is based on the dynamic Lagrangian (\ref{lag}).

To obtain the gravitational equations
related to the Lagrangian (\ref{lag}) one needs to vary it with respect to $h^a$. Using the property, ${\delta \lag (\bar g+h,\,\bar
\Phi+\phi)}/{\delta \bar g^a} = {\delta \lag (\bar g+h,\,\bar
\Phi+\phi)}/{\delta h^a}$,  we present them in the form:
 \be
\frac{\delta \lag^{dyn}}{\delta h^a} = \frac{\delta }{\delta \bar
g^a}\l[\lag(\bar g+h,\,\bar \Phi+\phi) - \bar \lag\r] = 0\, .
 \m{PERTeqs1}
 \ee
From here it is clear that the field equations for $h^a$ are {\em equivalent} to the gravitational equations in the standard form
(\ref{ddd}) if the background equations (\ref{bbb}) hold.

Now we define a generalized metric energy-momentum as
 \be
{\bm t}_a \equiv \frac{\delta \lag^{dyn}}{\delta \bar g^a} \equiv
\frac{\delta \lag^{dyn}}{\delta h^a} -\frac{\delta }{\delta \bar
g^a}\l(h^b \frac{\delta \bar \lag}{\delta \bar g^b} + \phi^A
\frac{\delta \bar \lag}{\delta \bar \Phi^A}\r)\,.
 \m{PERTcurrent}
 \ee
Note that the background equations should not  be
taken into account before variation of $\lag^{dyn}$ with respect to
$\bar g^a$ and $\bar \Phi^A$. Then, combining the definition (\ref{PERTcurrent}) with the dynamical equations
(\ref{PERTeqs1}) and contracting it with $2\k\di \Bar g^a/\Bar
g^{\mu\nu}$,  one obtains another form of the equations
(\ref{PERTeqs1}):
  \be
{\cal G}^{L}_{\mu\nu} + {\cal F}^{L}_{\mu\nu} = \k{\bm t}_{\mu\nu}\,,
 \m{PERTmunu}
 \ee
 which are, of course,  equivalent to the equations
(\ref{ddd}) if the background equations (\ref{bbb}) hold. The  linear operators on the
left hand side of (\ref{PERTmunu}) are defined by
 \bea
 {\cal G}^{L}_{\mu\nu} =
 \frac{\delta }{\delta \bar g^{\mu\nu}} h^a
\frac{\delta \bar \lag_{g}}{\delta \bar { g}^{a}}\,,
 \m{GL-q}\\
  {\cal F}^{L}_{\mu\nu} =
 -2\k\frac{\delta }{\delta \bar g^{\mu\nu}}\l(h^a
\frac{\delta \bar \lag_{m}}{\delta \bar { g}^{a}} + \phi
\frac{\delta \bar \lag_{m}}{\delta \bar \Phi} \r)\,.
 \m{PhiL-q}
 \eea
The right hand side in (\ref{PERTmunu}) is a total symmetric
(metric) energy-momentum tensor density for the fields (perturbations) $h^a$ and $\phi^A$ and
is defined by the standard varying of the Lagrangian (\ref{lag})
 \be
 {\bm t}_{\mu\nu} = 2\frac{\delta\lag^{dyn}}{\delta \bar
 g^{\mu\nu}} = {\bm t}^g_{\mu\nu} + {\bm t}^m_{\mu\nu}\, .
 \m{em-q}
 \ee
Here, ${\bm t}^g_{\mu\nu}$  is the energy-momentum related to a pure gravitational part of the Lagrangian (\ref{lag}); ${\bm t}^m_{\mu\nu}$ is the energy-momentum of matter fields $\phi^A$ in (\ref{lag}) interacting with the gravitational field $h^a$.

Now let us discuss a concrete choice from the set of variables in (\ref{(1)}).
Of course, this leads to different decompositions (\ref{g-Dec})
and different definitions of $h^a$. Taking into account the gravitational
background equations (\ref{bbb}) one can rewrite the left hand side
of the equations
 (\ref{PERTmunu}) as
  \be
 {\cal G}^{L}_{\mu\nu}+ {\cal F}^{L}_{\mu\nu} =
 -2\k\frac{\delta }{\delta \bar g^{\mu\nu}}\l(\goh^{\alf\beta}_{a}
\frac{\delta}{\delta \bar {\gog}^{\alf\beta}}\l(-\frac{1}{2\k}{
\bar \lag_{g}}+  \bar \lag_{m}\r) +\phi^A \frac{\delta \bar
\lag_{m}}{\delta \bar \Phi^A} \r)
 \m{GLPhiL-q}
 \ee
with independent (unified) gravitational variables
 \be
 \goh^{\mu\nu}_{a} =
 h^a {{ \di \bar {\gog}^{\mu\nu}} \over {\di \bar g^a}}\, ,
 \m{B40}
 \ee
 where $\gog^{\mu\nu} = \sqrt{-g}g^{\mu\nu}$.
The variables $\goh^{\mu\nu}_{a}$ in (\ref{B40}) differ one from
another beginning from the second order in perturbations. Of course, this
difference is incorporated into the left hand side of
(\ref{PERTmunu}), see (\ref{GLPhiL-q}). The right hand
side of (\ref{PERTmunu}) incorporates this difference as well. As a rule, such a
difference is not important for calculating conserved quantities for
static solutions; it also does not influence, e.g., in calculations
in quantum gravity \cite{B-Deser}. However, a difference in the second
order becomes important, for example, in a real calculation for a radiating
isolated system in 4D GR. It turns out that {\em only} the choice of the metric perturbations
 $
 h^a = \gog^{\mu\nu} - \bar {\gog}^{\mu\nu} = \goh^{\mu\nu}\,
  $
gives, see \cite{PK}, the standard Bondi-Sachs momentum
\cite{BMS}. All the other decompositions, including the popular one
$ h^a = g_{\mu\nu} - \bar g_{\mu\nu}\equiv \varkappa_{\mu\nu}$ (see, for example, \cite{AbbottDeser82}), do not.

\section{Covariantized Klein-Noether identities}
 \m{NIdentities}
\setcounter{equation}{0}

In this section, we present the necessary identities and their
important consequences. Following the technique developed in \cite{Petrov2009,Petrov2011}, see also the book \cite{Petrov+_2017},
we start from the Klein-Noether identities for an arbitrary set of covariant fields, see book \cite{Mitzk}.  Then, introducing an external metric field, we covariantize these identities.

Let a set of physical fields, $\psi^A$, be described by the Lagrangian:
 \be
\lag = \lag (\psi^A; \psi^A{}_{,\alf}; \psi^A{}_{,\alf\beta}) \m{lagQ}
 \ee
depending on partial derivatives up to the second order. Because the Lagrangian
(\ref{lagQ}) is a scalar density of the weight +1 it satisfies the main Noether identity:
 \be
 {\pounds}_\xi {\lag} + (\xi^\alf {\lag})_{,\alf} \equiv 0\,.
 \m{LieLag}
 \ee
After identity transformations it can be represented in the form:
 \be
 \di_\alf \l[{\cal U}_\sig{}^\alf\xi^\sig + {\cal M}_{\sig}{}^{\alf\tau}\di_\tau \xi^\sig + {\cal N}_\sig{}^{\alf\tau\beta}\di_{\beta\tau} \xi^\sig\r] \equiv 0\,.
 \m{(+2+)}
\ee
In (\ref{(+2+)}), the coefficients are defined by the Lagrangian (\ref{lagQ}) without ambiguities in an unique way:
 \bea
&& {\cal U}_\sig{}^\alf  =  \lag
\delta^\alf_\sig +
 {{\delta \lag} \over {\delta \psi_B}} \l.\psi_B\r|^\alf_\sig -
 \l[{{\di \lag} \over {\di \psi_{B,\alf}}} -
 \di_\beta
\l({{\di \lag} \over {\di \psi_{B,\alf\beta}}}\r) \r] \di_\sig
\psi_{B}  -  {{\di \lag} \over {\di \psi_{B,\alf\beta}}}
\di_{\sig\beta} \psi_{B}\, , \m{(+3+)} \\
 && {\cal M}_\sig{}^{\alf\tau}  =
 \l[{{\di \lag} \over {\di \psi_{B,\alf}}} -
 \di_\beta \l({{\di \lag} \over {\di \psi_{B,\alf\beta}}}\r)\r]
 \l.\psi_{B}\r|^\tau_\sig -
 {{\di \lag} \over {\di \psi_{B,\alf\tau}}}
\di_\sig \psi_B +
 {{\di \lag} \over {\di \psi_{B,\alf\beta}}}
 \di_\beta (\l.\psi_{B}\r|^\tau_\sig)\, ,
\m{(+4+)}\\
&& {\cal N}_\sig{}^{\alf\tau\beta} = \half \l[{{\di \lag} \over
{\di\psi_{B,\alf\beta}}}
 \l.\psi_{B}\r|^\tau_\sig +
 {{\di \lag} \over {\di \psi_{B,\alf\tau}}}
 \l.\psi_{B}\r|^\beta_\sig\r].
\m{(+5+)}
 \eea
Because $\di_{\beta\tau}\equiv \di_\beta\di_\tau$ in (\ref{(+2+)}) is symmetrical in $\beta$ and $\tau$ the same symmetry is reflected in coefficients:
${\cal N}_\sig{}^{\alf\tau\beta} = {\cal N}_\sig{}^{\alf\beta\tau}$.

Opening the identity (\ref{(+2+)}), given that $\xi^\sig$, $
\di_{\alf}\xi^\sig$, $ \di_{\beta\alf}\xi^\sig$ and
$\di_{\gamma\beta\alf} \xi^\sig$ are arbitrary at every world point,
we equate to zero the coefficients associated with them and obtain
the system of the Klein-Noether identities:
 \bea
 &{}& \di_\alf  {\cal U} _\sig{}^\alf  \equiv 0, \m{(+9+1A)}\\
&{}&    {\cal U}_\sig{}^\alf + \di_\lam {\cal M}_{\sig}{}^{\lam \alf}
\equiv 0,
 \m{(+9+2A)}\\ &{}&
 {\cal M}_{\sig}{}^{(\alf\beta)}+
\di_\lam  {\cal N}_{\sig}{}^{\lam(\alf\beta)} \equiv 0, \m{(+9+3A)}\\
&{}&
 {\cal N}^{(\alf\beta\gamma)}_\sig \equiv 0.
 \m{(+9+4A)}
 \eea
 These identities are not independent. Indeed, after differentiating (\ref{(+9+2A)}) and using (\ref{(+9+3A)}) and (\ref{(+9+4A)}) one easily finds (\ref{(+9+1A)}).

One can see that expressions (\ref{(+3+)}) - (\ref{(+5+)}) and
the identities (\ref{(+9+1A)}) - (\ref{(+9+4A)}) are not covariant.
On the other hand, the expression in  (\ref{(+2+)}) is  covariant as whole, since it
is a scalar density; the expression under the divergence in
(\ref{(+2+)}) is a vector density. This signals that the above expressions and the identities can be covariantized. We achieve this in the following way, see \cite{Petrov_Lompay_2013}.
Let us replace partial derivatives of $\xi^\sig$ in  (\ref{(+2+)}) by covariant ones, making the use of the expression
$\di_\rho\xi^\sig = \nabla_\rho\xi^\sig - \l.\xi^\sig\r|^\alf_\beta
\Gamma^\beta_{\rho\alf}$. Here, the Christoffel symbols $\Gamma^\beta_{\rho\alf}$ and, consequently, the covariant derivative $\nabla_\rho$ are compatible with $g_{\mu\nu}$. At the moment, $g_{\mu\nu}$ is included in expressions as an external metric only. Note that $g_{\mu\nu}$ can be included in the set $\psi^A$, although this is not necessary.  The identity (\ref{(+2+)}) is now
rewritten as
 \be
 \nabla_\alf\l[{\bm u}_\sig{}^\alf\xi^\sig + {\bm m}_\sig{}^{\alf\tau}\nabla_\tau \xi^\sig + {\bm n}_\sig{}^{\alf\tau\beta} \nabla_{\beta\tau}
\xi^\sig\r]\equiv 0\,,
 \m{secondID}
 \ee
where $\nabla_{\beta\tau}\equiv \nabla_{\beta} \nabla_{\tau}$ and
 \bea
 {\bm u}_\lam{}^\alf &=& {\cal U}_\lam{}^\alf - {\cal M}_\sig{}^{\alf\tau}\Gamma^\sig_{\lam\tau}+ {\cal  N}_\sig{}^{\alf\tau\rho} (\Gamma^\sig_{\tau\pi}\Gamma^\pi_{\lam\rho}
 - \di_\rho\Gamma^\sig_{\lam\tau})\,,\m{Uu}\\
 {\bm m}_\lam{}^{\alf\tau} &=& {\cal M}_\lam{}^{\alf\tau} +
 {\cal N}_\lam{}^{\alf\sig\rho}\Gamma^\tau_{\sig\rho} -
 2{\cal N}_\sig{}^{\alf\tau\rho}\Gamma^\sig_{\lam\rho}\,,
 \m{Mm}\\
 {\bm n}_\lam{}^{\alf\tau\rho} &=& {\cal N}_\lam{}^{\alf\tau\rho}\,.
 \m{Nn}
 \eea
 One can show explicitly that, indeed, these coefficients are covariant, see \cite{Petrov_Lompay_2013}. Note that in \cite{Petrov_Lompay_2013} we have shown that there are different ways to define coefficients in (\ref{Uu}), (\ref{Mm}) and (\ref{Nn}). Here, however, we use the form  (\ref{Uu}), (\ref{Mm}) and (\ref{Nn}) only.

 The identity (\ref{secondID}) can be rewritten in the form of the
differential conservation law:
 \be
 \nabla_\alf {\bm i}^\alf \equiv
 \di_\alf {\bm i}^\alf \equiv 0\,,
 \m{(+6+)}
 \ee
where the current is rewritten as
 \bea
 {\bm i}^\alf  &=& - \l[({\bm u}_\sig{}^\alf + {\bm n}_\lam{}^{\alf\beta\gamma} R^\lam_{~\beta\gamma\sig})\xi^\sig + {\bm m}^{\rho\alf\beta}\di_{[\beta}\xi_{\rho]} +
 {\bm z}^{\alf}\r]\,,
 \m{(+7+)}\\
  {\bm z}^{\alf} &=& {\bm m}^{\sig\alf\beta}\zeta_{\sig\beta}+ {\bm n}^{\rho\alf\beta\gamma}
\l(2 \nabla_{\gamma}\zeta_{\beta\rho} - \nabla_\rho \zeta_{\beta\gamma}\r); \qquad 2\zeta_{\rho\sigma} \equiv - {\pounds}_\xi g_{\rho\sigma} =
2\nabla_{(\rho}\xi_{\sigma)}\,.
 \m{(+8+)}
 \eea
Thus, $z$-term disappears, ${\bm z}^{\alf} = 0$, if $\xi^\mu = \bar \xi^\mu$ is a Killing vector of a
spacetime with the metric $g_{\mu\nu}$. Then the current (\ref{(+7+)}) is determined by the
energy-momentum $(u + nR)$-term and the spin $m$-term only.

Now, opening the identity (\ref{(+6+)}) and, equating independently
to zero the coefficients at $\xi^\sig$,  $\nabla_{\alf} \xi^\sig$,
$\nabla_{(\beta\alf)} \xi^\sig$ and $\nabla_{(\gamma\beta\alf)} \xi^\sig$
we get a system of identities that is equivalent to the system (\ref{(+9+1A)}) - (\ref{(+9+4A)})  and reformulates it as
 \bea
 &{}& \nabla_\alf  {\bm u}_\sig{}^\alf + \half
 {\bm m}_\lam{}^{\alf\rho} R^{~\lam}_{\sig~\rho\alf}
 +{\textstyle{1\over 3}} {\bm n}_\lam{}^{\alf\rho\gamma}
\nabla_\gamma R^{~\lam}_{\sig~\rho\alf}
  \equiv 0, \m{(+9+1)}\\
&{}&    {\bm u}_\sig{}^\alf + \nabla_\lam {\bm m}_{\sig}{}^{\lam \alf} +
{\bm n}_\lam{}^{\tau\alf\rho} R^{~\lam}_{\sig~\rho\tau} +{\textstyle{2\over 3}} {\bm n}_{\sig}{}^{\lam\tau\rho}R^{\alf}_{~\tau\rho\lam} \equiv 0,
 \m{(+9+2)}\\ &{}&
 {\bm m}_{\sig}{}^{(\alf\beta)}+
\nabla_\lam  {\bm n}_{\sig}{}^{\lam(\alf\beta)} \equiv 0, \m{(+9+3)}\\
&{}&
 {\bm n}^{(\alf\beta\gamma)}_\sig \equiv 0.
 \m{(+9+4)}
 \eea
 These identities are also not independent. After covariant differentiating (\ref{(+9+2)}) and using (\ref{(+9+3)}) and (\ref{(+9+4)}) one easily finds (\ref{(+9+1)}).
Since (\ref{(+6+)}) is identically satisfied, the current
(\ref{(+7+)}) must be a divergence of a superpotential $ {\bm i}^{\alf\beta}$, antisymmetrical tensor density, for which
$\di_{\beta\alf} {\bm i}^{\alf\beta} \equiv 0$, that is
 \be
 {\bm i}^\alf \equiv \nabla_{\beta} {\bm i}^{\alf\beta} \equiv
\di_{\beta}   {\bm i}^{\alf\beta}.
 \m{(+10+)}
 \ee
Indeed,  substituting ${\bm u}_\sig{}^\alf$  from (\ref{(+9+2)})
 into the current (\ref{(+7+)}), using (\ref{(+9+3)}) and algebraic properties of
${\bm n}_\sig{}^{\alf\beta\gamma}$ and $R^{\alf}_{~\beta\rho\sig}$ we
reconstruct (\ref{(+7+)}) to the form (\ref{(+10+)}), where  the
superpotential acquires the form:
 \be
 {\bm i}^{\alf\beta}  = \l({\textstyle{2\over 3}}
 \nabla_\lam  {\bm n}_{\sig}{}^{[\alf\beta]\lam}  - {\bm m}_{\sig}{}^{[\alf\beta]}\r)\xi^\sig   -
 {\textstyle{4\over 3}} {\bm n}_{\sig}{}^{[\alf\beta]\lam}
 \nabla_\lam  \xi^\sig.
 \m {(+11+)}
 \ee
It is explicitly antisymmetric in $\alf$ and $\beta$. Of course,
the differential conservation law
(\ref{(+6+)}) follows from (\ref{(+10+)}).

For our goals, to construct conserved quantities for perturbations in the Lovelock theory, we fix the Lagrangian (\ref{lagQ}) in a more concrete form:
 \be
 \lag = \lag(h^a; g_{\pi\sig};
 R^\alf{}_{\mu\beta\nu})\,.
 \m{LagR}
 \ee
 Thus, the set of fields in  (\ref{lagQ}) is now $\psi^A = \{h^a,\, g_{\pi\sig} \}$. The Lagrangian  (\ref{LagR}) is an arbitrary enough smooth algebraic function of $h^a$, $g_{\pi\sig}$ and the Riemannian tensor $R^\alf{}_{\mu\beta\nu}$. Note that derivatives of $g_{\pi\sig}$ are
included in $R^\alf{}_{\mu\beta\nu}$ only.

Now, let us define an important quantity
 \be
 {\bm \omega}^{\rho\lam|\mu\nu}  =   \frac{\di \lag}{\di
g_{\rho\lam,\mu\nu}}\,
 \m{NL1}
 \ee
with the evident symmetries
 \be
 {\bm \omega}^{\rho\lam|\mu\nu}  = {\bm \omega}^{\lam\rho|\mu\nu}  =
 {\bm \omega}^{\rho\lam|\nu\mu}\,.
  \m{omega}
 \ee
Because the Riemannian tensor is linear in derivatives of
$g_{\pi\sig}$ we conclude that the quantity (\ref{NL1}) is covariant automatically.
 Then, making the use of the definitions for the coefficients (\ref{(+3+)}) - (\ref{(+5+)})
 and (\ref{Uu}) - (\ref{Nn}),
 the identities  (\ref{(+9+4A)}) and  (\ref{(+9+4)}), the
quantity (\ref{NL1}) and its symmetries (\ref{omega}) we can rewrite (\ref{Uu}) - (\ref{Nn}) for the Lagrangian (\ref{LagR}) as
  \bea
{\bm u}_{\sig}{}^{\mu} & =& \lag\delta_\sigma^\mu +\frac{\delta
\lag}{\delta \psi^A}\l.\psi^A \r|^\mu_\sigma   - {\bm \omega}^{\lambda\mu|\rho\nu} R_{\lambda\rho\nu\sigma}
\, ,\m{uL}\\
 {\bm m}^{\rho\mu\nu}  &=& 2 \nabla_\lam
 {\bm \omega}^{\rho\nu|\mu\lam}; \qquad   {\bm m}^{\rho\mu\nu} =
 {\bm m}^{\nu\mu\rho}\, ,\m{mL} \\
 {\bm n}^{\rho\lam\mu\nu} & =& {\bm \omega}^{\rho\lam|\mu\nu};\qquad\qquad{\bm \omega}^{\rho\lam|\mu\nu} = {\bm \omega}^{\mu\nu|\rho\lambda}\,.
 \m{nL}
 \eea
One can see that all the expressions (\ref{uL}) - (\ref{nL}) are covariant, and, of course, they satisfy the identities (\ref{(+9+2)}) -
(\ref{(+9+4)}).

Now, making the use of (\ref{uL}) - (\ref{nL}) we represent
the current (\ref{(+7+)}) - (\ref{(+8+)}) and the superpotential
(\ref{(+11+)}) for the Lagrangian (\ref{LagR}) as
  \bea
{\bm i}^\mu &=& -\l(\lag\xi^\mu + \frac{\delta \lag}{\delta
\psi^A}\l.\psi^A \r|^\mu_\sigma \xi^\sig + {\bm z}^\mu\r)\,,
 \label{Jmu}\\
  {\bm z}^\mu &= & 2\zeta_{\rho\lam}\nabla_\nu
 {\bm \omega}^{\rho\lam|\mu\nu}- 2{\bm \omega}^{\rho\lam|\mu\nu}
  \nabla_\nu \zeta_{\rho\lam} \,.\label{ZDmu}\\
{\bm i}^{\mu\nu} & =&{\textstyle{4\over 3}}\l(
 2\xi^\sig \nabla_\lam  {\bm \omega}_{\sig}{}^{[\mu|\nu]\lam}   -
{\bm \omega}_{\sig}{}^{[\mu|\nu]\lam}
 \nabla_\lam  \xi^\sig\r)\,.
 \m{(+16+A)}
 \eea
Let us note remarkable properties of these quantities. First, due to the symmetry in (\ref{mL}) $m$-term disappears from the
current, compare (\ref{(+7+)}) with (\ref{Jmu}). Second, the expression for the superpotential
  (\ref{(+16+A)}) depends essentially  on the quantity (\ref{NL1}) {\em only}.

Expressions for currents ${\bm i}^\mu$ and superpotentials ${\bm i}^{\mu\nu}$ in (\ref{Jmu}) - (\ref{(+16+A)}) constructed in the case of absence of
$h^a$ in the Lagrangian (\ref{LagR}) can be considered as an intermediate result. In this case, thus, we consider unperturbed theories that can be not only the Lovelock theory, but  theories quadratic in curvature (see, for example, \cite{DT1,DT2}), or $f(R)$ gravities (see \cite{Sotiriou}), etc. Essentially, it can be an arbitrary theory with the Lagrangian $\lag = \lag(g_{\pi\sig}; R^\alf{}_{\mu\beta\nu})$ that is an arbitrary enough smooth algebraic function of  $g_{\pi\sig}$ and $R^\alf{}_{\mu\beta\nu}$.
Then the superpotential (\ref{(+16+A)}) can be interpreted as a generalization of the well known Komar
superpotential \cite{Komar59}. Indeed, let the Lagrangian (\ref{LagR}) be the Hilbert Lagrangian
\be
 \lag_h = -\frac{1}{2\k}\sqrt{-g} R
\,.
 \m{LagH}
 \ee
 Then the quantity (\ref{NL1}) acquires the form:
\be
{\bm \omega}^{\rho\lam|\mu\nu}_h  = -\frac{1}{4\k}\l(g^{\rho\mu}g^{\nu\lam} + g^{\rho\nu}g^{\mu\lam} - 2g^{\rho\lam}g^{\mu\nu} \r)\,,
 \m{NL1H}
 \ee
 and the superpotential (\ref{(+16+A)}) becomes the famous Komar superpotential \cite{Komar59}:
 \be
 {\bm i}^{\mu\nu}_h = \frac{\sqrt{-g}}{\k}\nabla^{[\mu}\xi^{\nu]}\,.
 \m{Komar}
 \ee
 See also \cite{Lompay_Petrov_2013a,Lompay_Petrov_2013b} for a generalization of the Komar superpotential in metric theories with torsion.

\section{Currents and superpotentials of general form}
 \m{ConservedCurrents}
 \setcounter{equation}{0}

In the present section, we use the identities of the above section to construct currents and superpotentials in metric theories in the framework of the field-theoretical formulation. Our approach is based on exploiting the structure of the Lagrangian $\lag^{dyn}$ defined in (\ref{lag}). Let us consider the part of $\lag^{dyn}$,
\be
\lag_1 = -\frac{1}{2\k}h^{a} \frac{\delta \bar
\lag_{g}}{\delta \bar {g}^a}\, ,
 \m{Lag-1}
 \ee
that now plays a role of an {\em auxiliary} Lagrangian (\ref{LagR}), and where we can set $\psi^A = \{h^a,\,\bar g_{\pi\sig} \}$.
 Index "$_1$" is used because Lagrangian (\ref{Lag-1}) is of the first order in  $h^{a}$ in expansion of $\lag_{g}$. To apply the technique of the previous section to the Lagrangian (\ref{Lag-1}), we assume that $\lag_1 = \lag_1(h^a; \bar g_{\pi\sig};
\bar R^\alf{}_{\mu\beta\nu})$. Then, because $\lag_1$ in (\ref{Lag-1}) is proportional to the Lagrange derivative of $\bar \lag_{g}$, the gravitational part of the Lagrangian (\ref{lag-g}) can present the Lovelock theory only, see \cite{Lovelock} and section \ref{Lovelock}. However, in the present section, we do not use concrete expressions of the Lovelock gravity and obtain identically conserved quantities related to the Lagrangian (\ref{Lag-1}) in a general form $\lag_1 = \lag_1(h^a; \bar g_{\pi\sig};
\bar R^\alf{}_{\mu\beta\nu})$ only. Because the Lagrangian (\ref{Lag-1}) induces the construction of the linear operator (\ref{GL-q}) in (\ref{PERTmunu}), making the use of the field equations (\ref{PERTmunu}), we transform the {\em identically} conserved quantities to {\em physically} conserved quantities.

The explicit expression for the linear operator ${\cal G}^{L}_{\mu\nu}$ in (\ref{PERTmunu}) defined in (\ref{GL-q}) is quite important in numerous applications. Let us derive it. For the Lagrangian (\ref{Lag-1}) that is of the type (\ref{LagR}) and for the related expressions (\ref{uL}) - (\ref{nL}), one finds from the identity (\ref{(+9+2)}):
 \be
{\cal G}_\sigma^{L\mu} =- \frac{1}{2}\frac{\delta \bar \lag_{g}}{\delta
\bar {g}^{a}}\l.\l(h^a\delta^\mu_\sig + h^a\r|^\mu_\sig\r) +2\k \l(
\bar\nabla_{\rho\lam}  {\bm \omega}_{1\sig}{}^{\mu|\rho\lam} +
 {\bm \omega}_1^{\mu\tau|\rho\lam} \bar R_{\sig\lam\tau\rho} + \frac{1}{3} {\bm \omega}_{1\sig}{}^{\lam|\tau\rho} \bar R^{\mu}{}_{\tau\rho\lam} \r)\, .
 \m{Linear-Gen}
 \ee
Note that this formula can be obtained following the method in \cite{Rund},
however, the use of the identities (\ref{(+9+1)}) - (\ref{(+9+4)}) is more clear and simple.

Next, substituting the expression for the Lagrangian (\ref{Lag-1}) into
the current expression (\ref{Jmu}), one obtains
  \be
{\bm i}_1^\mu = {\bm \theta}_\sigma{}^\mu\xi^\sigma  -
{\bm z}^\mu_{1} \,,
 \label{Jmu1}
 \ee
 where, as usual, the coefficient ${\bm \theta}_\sigma{}^\mu$ at $\xi^\sigma$ is interpreted as the energy-momentum,
  \be
{\bm \theta}_\sigma{}^\mu =\frac{1}{\k} \l({\cal G}_\sigma^{L\mu} + \frac{1}{2}\frac{\delta \bar \lag_{g}}{\delta
\bar {g}^{a}}\l.\l(h^a\delta^\mu_\sig + h^a\r|^\mu_\sig\r) \r) \,,
 \label{EM1}
 \ee
 and
 z-term ${\bm z}^\mu_{1}$  has exactly the form
(\ref{ZDmu}) with ${\bm \omega^{\rho\lam|\mu\nu}_1}$ defined in (\ref{NL1}) and related to $\lag_1$.  Combining the last expression with  (\ref{Linear-Gen}), one finds a quite simple formula for the energy-momentum:
 \be
{\bm \theta}_\sigma{}^\mu =2\l(
\bar\nabla_{\rho\lam}  {\bm \omega}_{1\sig}{}^{\mu|\rho\lam} +
 {\bm \omega}_1^{\mu\tau|\rho\lam} \bar R_{\sig\lam\tau\rho} + \frac{1}{3} {\bm \omega}_{1\sig}{}^{\lam|\tau\rho} \bar R^{\mu}{}_{\tau\rho\lam} \r) \,.
 \label{EM1A}
 \ee
At the moment, formally the energy-momentum (\ref{EM1A}) is related to the auxiliary Lagrangian (\ref{Lag-1}), not more. One can remark also the nice property of the expression (\ref{EM1A}) that, being quite general, depends {\em essentially } on the quantity ${\bm \omega}_{1}^{\sig\lam|\tau\rho}$ only.

Of course, the current (\ref{Jmu1}) is identically conserved:
 \be
 \bar\nabla_\mu {\bm i}_1^\mu \equiv \di_\mu {\bm i}_1^\mu
\equiv 0\, .\m{BDi-1}
 \ee
 Then, the current in this identity, with making the use of the Klein-Noether identities, can be rewritten with the use of a superpotential
 \be
 {\bm i}_1^\mu \equiv \bar\nabla_\nu {\bm i}_1^{\mu\nu} \equiv \di_\nu {\bm i}_1^{\mu\nu}
\, ,\m{BDi-sup}
 \ee
 where ${\bm i}_1^{\mu\nu}$ has exactly the form
(\ref{(+16+A)}) with ${\bm \omega^{\rho\lam|\mu\nu}_1}$ related to $\lag_1$.

However, both (\ref{BDi-1}) and (\ref{BDi-sup}) are the identities only. To make them physically sensible  conservation laws one has to use the field equations. Thus, after using (\ref{PERTmunu}) the energy-momentum (\ref{EM1}) transforms to
\be
{\bm \theta}_{\mu\nu} \goto {\bm \tau}_{\mu\nu} ={\bm t}^g_{\mu\nu} + {\bm t}^m_{\mu\nu} - \frac{1}{\k} {\cal F}^{L}_{\mu\nu} + \frac{1}{2\k}\frac{\delta \bar \lag_{g}}{\delta
\bar {g}^{a}}\l.\l(h^a  \bar g_{\mu\nu} + h^a\r|_\mu^\sig\bar g_{\sig\nu}\r) \,.
 \label{EM1+}
 \ee
Let us interpret this expression. The first term is the energy-momentum for a free gravitational field related to the gravitational part of the Lagrangian (\ref{lag}); the second term is the energy-momentum for matter fields related to the matter part of the Lagrangian (\ref{lag}); the last term describes interaction of the gravitational field $h^a$ with a curved background described by the metric $\bar g_{\mu\nu}$. However, the role of the third term is not clear. Let us clarify it. Using the definitions (\ref{PhiL-q}) and (\ref{em-q}), uniting the second and third terms, and taking into account (\ref{ddd}) and (\ref{bbb}), one transforms (\ref{EM1+}) to
\be
{\bm \tau}_{\mu\nu} ={\bm t}^g_{\mu\nu} + \delta{\cal T}_{\mu\nu}+ \frac{1}{2\k}\frac{\delta \bar \lag_{g}}{\delta
\bar {g}^{a}}\l.\l(h^a  \bar g_{\mu\nu} + h^a\r|_\mu^\sig\bar g_{\sig\nu}\r)  \,,
 \label{EM1++A}
 \ee
 where $\delta {\cal T}_{\mu\nu} = {\cal T}_{\mu\nu} -  \bar{\cal T}_{\mu\nu}$ describes a perturbation of the matter energy-momentum of the gravity theory in (\ref{ddd}) with respect to the background one in (\ref{bbb}). Of course, if we examine a concrete solution to the field equations
(\ref{PERTmunu}), we can use the energy-momentum (\ref{EM1A}),  instead of (\ref{EM1+}) or (\ref{EM1++A}). Thus, it is
\be
{\bm \tau}^{\mu\nu} = 2\l.\l(
\bar\nabla_{\rho\lam}  {\bm \omega}_{1}^{\mu\nu|\rho\lam} +
 {\bm \omega}_1^{\mu\tau|\rho\lam} \bar R^{\nu}{}_{\lam\tau\rho} + \frac{1}{3} {\bm \omega}_{1}^{\nu\lam|\tau\rho} \bar R^{\mu}{}_{\tau\rho\lam} \r)\r|_{(\ref{PERTmunu})} \,.
\m{EM1+++}
\ee

After that let us turn to the current (\ref{Jmu1}) and transform it to
 \be
{\bm i}_1^\mu \goto {\cal I}^\mu  =   {\bm \tau}^{\mu\nu}\xi_\nu - {\bm z}^\mu_{1}\,.
 \m{current}
 \ee
Then the identity (\ref{BDi-1}) transforms to the physically sensible conservation law:
 \be
\bar\nabla_\mu  {\cal I}^\mu =\di_\mu  {\cal I}^\mu = 0\,.
 \m{nablacurrent+}
 \ee
 At last, substituting potentials of a concrete solution into the superpotential expression (\ref{BDi-sup}) we denote it as
 \be
{\bm i}_1^{\mu\nu} \goto {\cal I}^{\mu\nu}  \,.
 \m{current++}
 \ee
Then the identity (\ref{BDi-sup}) transforms to the physically sensible conservation law:
 \be
 {\cal I}^\mu = \bar\nabla_\nu {\cal I}^{\mu\nu} = \di_\nu {\cal I}^{\mu\nu}
\, .\m{BDi-sup+}
 \ee

All the above has been constructed for arbitrary curved backgrounds, even
non-vacuum ones. However,  the case of a vacuum background,
 \be
 \bar{\lag}_{m} = 0  \goto \qquad \frac{\delta \bar\lag_{g}}{\delta \bar {g}^{a}} =
\bar{\cal T}_{\mu\nu}=0,\qquad  {\cal F}^{L}_{\mu\nu}=0\,,
\m{vacuumB}
 \ee
is of special interest, and we describe this below. The field equations
(\ref{PERTmunu}) go to
\be
{\cal G}^{L}_{\mu\nu} = \k\l({\bm t}^g_{\mu\nu} + {\bm t}^m_{\mu\nu} \r)\,.
\m{EqsVacuum}
\ee
Note that it can be that ${\bm t}^m_{\mu\nu}\neq 0$ because, of course, matter can propagate on a vacuum background. Under the conditions (\ref{vacuumB}) the expression (\ref{Linear-Gen}) transforms to
\be
{\cal G}_\sigma^{L\mu} =2\k \l(
\bar\nabla_{\rho\lam}  {\bm \omega}_{1\sig}{}^{\mu|\rho\lam} +
 {\bm \omega}_1^{\mu\tau|\rho\lam} \bar R_{\sig\lam\tau\rho} + \frac{1}{3} {\bm \omega}_{1\sig}{}^{\lam|\tau\rho} \bar R^{\mu}{}_{\tau\rho\lam} \r)\,
  \m{Linear-Vac}
 \ee
 and
 the energy-momentum (\ref{EM1+}) acquires the simple form
\be
{\bm \tau}_{\mu\nu} = {\bm t}^g_{\mu\nu} + {\bm t}^m_{\mu\nu} \,.
\m{EM1++B}
\ee
In the case of a vacuum background, we highlight the nice property that the expression (\ref{Linear-Vac})  depends essentially on the quantity ${\bm \omega}_{1}^{\sig\lam|\tau\rho}$ only, defined as in (\ref{NL1}) for the Lagrangian $\lag_1$.
Next, in the case (\ref{vacuumB}), making use of the Killing vectors $\xi^\alf =\bar\xi^\alf$, the current (\ref{current}) transforms to the usual
form of a field theory:
  \be
{\cal I}^\mu = {\bm \tau}^{\mu\nu}\bar\xi_\nu\, .
 \m{current+1}
 \ee
Assuming {\em arbitrary} Killing vectors $\xi^\alf=\bar\xi^\alf$ in the identity
(\ref{BDi-1}), one obtains the identity
 \be
 \bar\nabla^\mu{\cal G}^{L}_{\mu\nu} \equiv 0
 \m{DGL-q}
 \ee
 under the conditions (\ref{vacuumB}). Then, from the field equations
(\ref{EqsVacuum}) one obtains the result,
 \be
 \bar\nabla_\nu {\bm\tau}^{\mu\nu}= 0,
 \m{CLfor-t}
 \ee
that is a differential conservation law for the total energy-momentum tensor density (\ref{EM1++B}).

\section{Lovelock theory. Arbitrary curved backgrounds}
 \m{Lovelock}
\setcounter{equation}{0}

In this section, we derive explicit expressions for the Lagrangian based field-theoretical reformulation of the Lovelock gravity including conserved currents and superpotentials. Let us consider Lagrangian (\ref{lag-g}) derived for the Lovelock theory:
 \be
 \lag(g^a,\Phi^A) = - \frac{1}{2\k}\lag_{\ell}(g^a) + \lag_{m}(g^a,\Phi^A)
 \,.
 \m{lag-ll}
 \ee
Here, $\k = 2\Omega_{D-2}G_D> 0$  with $G_D$ ---
$D$-dimensional Newton's constant, $\Omega_{D-2}$ is a square of $(D-2)$-dimensional sphere with unite radius. The Lovelock Lagrangian, $\lag_{\ell}(g^a)$, has been constructed under the following requirements \cite{Lovelock}: First, it has to be a Lagrangian of a covariant metric theory in $D$ dimensional spacetime; second, Euler-Lagrange equations following from varying $\lag_{\ell}(g^a)$ have to be differential equations of the second order, not more. The unique possibility to satisfy them is to construct a sum of polynomials in Riemannian tensor as
\begin{equation}
\lag_{\ell}(g^a)=\sqrt{-g}\sum_{p=0}^{m}\frac{\alf_p}{2^p} \delta _{\lbrack i_{1}i_{2}\cdots
i_{2p}]}^{[j_{1}j_{2}\cdots j_{2p}]}\,{R}_{j_{1}j_{2}}^{i_{1}i_{2}}\cdots {R}_{j_{2p-1}j_{2p}}^{i_{2p-1}i_{2p}}  \,,  \label{Lovelock-lag}
\end{equation}
where $\alf_p$ are coupling constants, $m=\l[(D-1)/2\r]$, and the totally-antisymmetric Kronecker delta is
\begin{equation}
\delta _{\left[ \mu _{1}\cdots \mu _{q}\right] }^{\left[ \nu _{1}\cdots \nu
_{q}\right] }:=\left\vert
\begin{array}{cccc}
\delta _{\mu _{1}}^{\nu _{1}} & \delta _{\mu _{1}}^{\nu _{2}} & \cdots &
\delta _{\mu _{1}}^{\nu _{q}} \\
\delta _{\mu _{2}}^{\nu _{1}} & \delta _{\mu _{2}}^{\nu _{2}} &  & \delta
_{\mu _{2}}^{\nu _{q}} \\
\vdots &  & \ddots &  \\
\delta _{\mu _{q}}^{\nu _{1}} & \delta _{\mu _{q}}^{\nu _{2}} & \cdots &
\delta _{\mu _{q}}^{\nu _{q}}%
\end{array}%
\right\vert \m{determinant}\,.
\end{equation}%

 Important properties of the Lagrangian (\ref{Lovelock-lag}) are as follow. Polynomial of zero's order, $p=0$, is related to a `bare' cosmological constant $\Lambda_0$ with $\alf_0 =-2\Lambda_0$; polynomial of the first order, $p=1$, is the Hilbert term with $\alf_1 = 1$, see (\ref{LagH}); polynomial of the second order, $p=2$, is the Gauss-Bonnet term with a coupling constant $\alf_2 $ undefined from the start, etc. The summation in (\ref{Lovelock-lag}) is restricted to $m=\l[(D-1)/2\r]$ for the next reasons. Consider, for example, dimensions $D=3,4$. In this case, $p=0$ and $p=1$ terms contribute to the Euler-Lagrange equations, whereas terms with $p \ge 2$ do not. For $D=5,6$, the terms with $p=0,1,2$ contribute to the Euler-Lagrange equations, whereas terms with $p \ge 3$ do not, etc. Although, of course, terms that do not contribute to equations of motion can be included in the Lagrangian as topological ones, and could play an important role, for example, in definition of conserved charges, see \cite{Olea_2005}.

Here and below, to concretize the decomposition (\ref{g-Dec}) we prefer to use a more simple and popular definition for gravitational variables $h^a$:
\be
\varkappa_{\mu\nu} = g_{\mu\nu} - \bar g_{\mu\nu}\,.
\m{h_a}
\ee
Therefore we derive the field equations related to (\ref{lag-ll}) by varying with respect to $g_{\rho\sig}$ with lower indices:
\begin{equation}
\frac{\delta\lag_{\ell}}{\delta g_{\rho\sig}}= \sqrt{-g}g^{\pi\rho}
\sum_{p=0}^{m}\frac{\alf_p}{2^{p+1}} \delta _{\lbrack \pi \nu_{1}\nu_{2}\cdots
\nu_{2p}]}^{[\sig \mu_{1}\mu_{2}\cdots \mu_{2p}]}\,{R}_{\mu_{1}\mu_{2}}^{\nu_{1}\nu_{2}}\cdots {R}_{\mu_{2p-1}\mu_{2p}}^{\nu_{2p-1}\nu_{2p}}  = - \kappa {\cal T}^{\rho\sig} \,.  \label{Lovelock-eqs+}
\end{equation}
As before, we do not concretize the matter sources in both, (\ref{lag-ll}) and (\ref{Lovelock-eqs+}), and below.

 Now we need to derive Lagrangian (\ref{Lag-1}) for the Lovelock theory. We use the background version of the Lovelock Lagrangian, $\bar\lag_{\ell}$, and the background version of the Lagrange derivative in (\ref{Lovelock-eqs+}). As a result,
the Lagrangian (\ref{Lag-1}) acquires the form:
\begin{equation}
\lag_{\ell 1}= - \frac{1}{2\k}\varkappa_{\rho\sig}\frac{\delta\bar\lag_{\ell}}{\delta \bar g_{\rho\sig}}=- \frac{\sqrt{-\bar g}}{2\k} \varkappa_\sig^\rho
\sum_{p=0}^{m}\frac{\alf_p}{2^{p+1}} \delta _{\lbrack \rho \nu_{1}\nu_{2}\cdots
\nu_{2p}]}^{[\sig \mu_{1}\mu_{2}\cdots \mu_{2p}]}\,{\bar R}_{\mu_{1}\mu_{2}}^{\nu_{1}\nu_{2}}\cdots {\bar R}_{\mu_{2p-1}\mu_{2p}}^{\nu_{2p-1}\nu_{2p}}   \,.  \label{lock-1}
\end{equation}
Another necessary and key quantity of the type (\ref{NL1}) calculated for the Lagrangian (\ref{lock-1}) is
\begin{equation}
{\bm \omega}^{\rho\lambda|\mu\nu}_{\ell 1}= - \frac{\sqrt{-\bar g}}{2\k} \varkappa^\alf_\beta
\sum_{p=1}^{m}\frac{p\alf_p}{2^{p+1}} \delta _{\lbrack \alf\phi\psi \nu_{3}\nu_{4}\cdots
\nu_{2p}]}^{[\beta\pi\sigma \mu_{3}\mu_{4}\cdots \mu_{2p}]}\,{\bar R}_{\mu_{3}\mu_{4}}^{\nu_{3}\nu_{4}}\cdots {\bar R}_{\mu_{2p-1}\mu_{2p}}^{\nu_{2p-1}\nu_{2p}}\bar g^{\phi\tau}\bar g^{\psi\kappa}D^{\rho\lambda\mu\nu}_{\pi\sigma\tau\kappa}  \,.  \label{NL1-lock}
\end{equation}
Here, the quantity
\be
D^{\rho\lambda\mu\nu}_{\pi\sigma\tau\kappa} = \half \l(\delta^\rho_\pi\delta^\lambda_\k +\delta^\rho_\k\delta^\lambda_\pi \r)\l(\delta^\mu_\sigma\delta^\nu_\tau +\delta^\mu_\tau\delta^\nu_\sigma \r)
\m{D}
\ee
has been obtained after differentiating the Riemannian tensor $\bar R_{\pi\sigma\tau\kappa}$ with respect to $\bar g_{\rho\lambda,\mu\nu}$ and using the index symmetry under the contraction.

In the framework of the Lovelock gravity, keeping in mind the quantity (\ref{NL1-lock}), first, we derive the linear operator (\ref{Linear-Gen}):
\be
{\cal G}^{\sigma\mu}_L =- \frac{1}{2}\bar g^{\sigma\mu}\frac{\delta \bar \lag_{\ell}}{\delta
\bar {g}_{\rho\tau}}\varkappa_{\rho\tau}+\frac{\delta \bar \lag_{\ell}}{\delta
\bar {g}_{\rho\sig}}\varkappa_{\rho}^{\mu} +2\k \l(
\bar\nabla_{\rho\lam}  {\bm \omega}_{\ell 1}^{\sig\mu|\rho\lam} +
 {\bm \omega}_{\ell 1}^{\mu\tau|\rho\lam} \bar R^{\sig}{}_{\lam\tau\rho} + \frac{1}{3} {\bm \omega}_{\ell 1}^{\sig\lam|\tau\rho} \bar R^{\mu}{}_{\tau\rho\lam} \r)\, ;
 \m{Linear-Genl}
 \ee
second, we derive the conserved current (\ref{current}):
 \be
{\cal I}^\mu_{\ell}  =   {\bm \tau}^{\sigma\mu}_{\ell}\xi_\sigma - {\bm z}^\mu_{\ell}\,
 \m{current-l}
 \ee
 with the energy-momentum (\ref{EM1+++}),
 \be
 {\bm \tau}^{\sigma\mu}_{\ell} = \l. 2\l(
\bar\nabla_{\rho\lam}  {\bm \omega}_{\ell 1}^{\sig\mu|\rho\lam} +
 {\bm \omega}_{\ell 1}^{\mu\tau|\rho\lam} \bar R^{\sig}{}_{\lam\tau\rho} + \frac{1}{3} {\bm \omega}_{\ell 1}^{\sig\lam|\tau\rho} \bar R^{\mu}{}_{\tau\rho\lam} \r)\r|_{(\ref{PERTmunu})}\,,
 \m{theta}
 \ee
 where $z$-term is
 \be
   {\bm z}^\mu_{\ell} =  2\bar\zeta_{\rho\lam}\bar\nabla_\nu
 {\bm \omega}_{\ell 1}^{\rho\lam|\mu\nu}- 2{\bm \omega}_{\ell 1}^{\rho\lam|\mu\nu}
  \bar\nabla_\nu \bar\zeta_{\rho\lam};\qquad 2\bar\zeta_{\rho\lam} \equiv -\Lix\bar g_{\rho\lam} = 2\bar\nabla_{(\rho}\xi_{\lam)}
  \,;\label{ZDmu-l}
 \ee
and, third, we derive the superpotential  (\ref{current++}):
\be
{\cal I}^{\mu\nu}_{\ell} = {\textstyle{4\over 3}}\l(
 2\xi_\sig \bar\nabla_\lam  {\bm \omega}_{\ell 1}^{\sig[\mu|\nu]\lam}  -
{\bm \omega}_{\ell 1}^{\sig[\mu|\nu]\lam}
\bar \nabla_\lam  \xi_\sig\r)\,.
 \m{Super-l}
 \ee
 We stress the remarkable property: For the Lovelock gravity both items of the conserved current (\ref{current-l}) and the superpotential (\ref{Super-l}) constructed for {\em arbitrary} perturbations on {\em arbitrary} curved backgrounds depend on the quantity (\ref{NL1-lock}) {\em only}. Recall also that the energy-momentum (\ref{theta}) is
 conserved for a vacuum background, see (\ref{CLfor-t}).

 \section{Lovelock theory. AdS backgrounds}
 \m{S-AdS-mass}
\setcounter{equation}{0}

One of the more popular solutions to equations of the Lovelock gravity is the global maximally symmetric spacetime with a negative constant curvature - anti-de Sitter (AdS) space. Therefore, it is very important to construct conserved quantities for arbitrary perturbations on such backgrounds. This is one of the tasks of the present section. The other task is to test the new formulae, calculating mass for a static black hole with the AdS asymptotic in the Lovelock gravity.

We consider the equations (\ref{Lovelock-eqs+}) as background equations under the vacuum condition (\ref{vacuumB}):
\begin{equation}
\frac{\delta\bar\lag_{\ell}}{\delta \bar g_{\rho\sig}}= \sqrt{-\bar g}\bar g^{\pi\rho}
\sum_{p=0}^{m}\frac{\alf_p}{2^{p+1}} \delta _{\lbrack \pi \nu_{1}\nu_{2}\cdots
\nu_{2p}]}^{[\sig \mu_{1}\mu_{2}\cdots \mu_{2p}]}\,{\bar R}_{\mu_{1}\mu_{2}}^{\nu_{1}\nu_{2}}\cdots {\bar R}_{\mu_{2p-1}\mu_{2p}}^{\nu_{2p-1}\nu_{2p}} = 0\,.  \label{Lovelock-eqs++}
\end{equation}
The AdS space with the Riemannian tensor
\be
\bar R^{\rho\lam}_{\mu\nu} = - \frac{1}{\ell^2_{eff}}\delta^{[\rho\lam]}_{[\mu\nu]}\,
\m{R_AdS}
\ee
satisfy the equations (\ref{Lovelock-eqs++}), for which effective cosmological constant is chosen as
\be
\Lambda_{eff} = -\frac{(D-1)(D-2)}{2\ell^2_{eff}}\,.
\m{Lambda_eff}
\ee
The  AdS radius $\ell_{eff}$ is defined by the value of constant curvature of the maximally symmetric spacetime solution (\ref{R_AdS}) to the vacuum equations (\ref{Lovelock-eqs++}). One easily finds that  $\ell_{eff}$ is to be a solution to the equation
\be
\l.V(x)\r|_{x =\ell_{eff}^{-2} } = \sum^{m}_{p=0}\frac{(D-3)!}{(D-2p-1)!}\alf_p(-1)^{p-1}\l(\ell_{eff}^{-2} \r)^p = 0\,.
\m{sum}
\ee
Thus, the AdS radius $\ell_{eff}$ (the same $\Lambda_{eff}$) defines a real (effective) length scale. On the other hand, the bare $\Lambda_0$ has no an analogical geometrical (physical) sense. Indeed, $\Lambda_0$ is related to the coupling constant $\alf_0$ only that is considered in (\ref{sum}) together with other $\alf_p$ only, there is no other sense for $\Lambda_0$. This treatment is close to the one in \cite{Giacomini+1,Giacomini+2}.

To describe a spacetime with the curvature tensor (\ref{R_AdS}) we can use the metric:
 \be
 d\bar s^2 = -\Bar fdt^2 + \frac{1}{\bar f}dr^2 +
 r^2\sum_{a,b}^{D-2}q_{ab}dx^adx^b\, ,
 \m{AdS_back}
 \ee
 where
 \be
\bar f(r) = 1+ \frac{r^2}{\ell^2_{eff}}\, ,
 \m{fBar}
 \ee
 and the last term in (\ref{AdS_back}) describes $(D-2)$-dimensional sphere of the
radius $r$, and $q_{ab}$ depends on coordinates on the sphere only. The Christoffel symbols corresponding (\ref{AdS_back}) are
 \be
 \bar\Gamma^1_{00} =  \frac{\bar f{ \bar f}'}{2} ,~~~
 \bar \Gamma^0_{10} = \frac{ \bar f'}{2 \bar f} ,~~~
 \bar \Gamma^1_{11} = -\frac{ \bar f'}{2 \bar f} ,~~~
 \bar \Gamma^a_{1b} =  \frac{1}{r}\delta^a_b ,~~~
 \Gamma^1_{ab} = -r \bar fq_{ab}\,,
 \m{ChristAdS}
 \ee
where $\di  \bar f/\di r = {\bar f}'$.

The unique key expression in constructing conserved quantities for perturbations (\ref{h_a}) on vacuum backgrounds in the Lovelock gravity is (\ref{NL1-lock}). Let us derive it taking into account the condition (\ref{R_AdS}) for the AdS background. From the beginning let us calculate the first term from the sum in (\ref{NL1-lock}) that corresponds to the Einstein part of the Lovelock Lagrangian. It is
\bea
&&{\bm \omega}^{\rho\lambda|\mu\nu}_{h1}= - \frac{\sqrt{-\bar g}}{8\k} \varkappa^\alf_\beta
\delta _{\lbrack \alf\phi\psi ]}^{[\beta\pi\sigma ]}\bar g^{\phi\tau}\bar g^{\psi\kappa}D^{\rho\lambda\mu\nu}_{\pi\sigma\tau\kappa} = \nonumber\\
&&- \frac{\sqrt{-\bar g}}{4\k}\l[\bar g^{\mu\nu}\varkappa^{\rho\lam} +\bar g^{\rho\lam}\varkappa^{\mu\nu} -\bar g^{\rho(\mu}\varkappa^{\nu)\lam} -\bar g^{\lam(\mu}\varkappa^{\nu)\rho} - \varkappa\l(\bar g^{\mu\nu}\bar g^{\rho\lam} - \bar g^{\rho(\mu}\bar g^{\nu)\lam} \r)\r] \,.  \label{NL1-H}
\eea
Now, keeping in mind this expression and the condition (\ref{R_AdS}), and making use of the relation for a contraction of indices in the Kronecker delta,
\begin{equation}
\delta _{\left[ \mu _{1}\cdots \mu _{2k}\mu _{2k+1}\cdots \mu _{2p}\right] }^{\left[ \nu
_{1}\cdots \nu _{2k}\nu _{2k+1}\cdots \nu _{2p}\right] }\,\delta _{\nu _{2k+1}}^{\mu
_{2k+1}}\cdots \delta _{\nu _{2p}}^{\mu _{2p}}=\frac{\left( D-2k\right) !}{%
\left( D-2p\right) !}\,\delta _{\left[ \mu _{1}\cdots \mu _{2k}\right] }^{%
\left[ \nu _{1}\cdots \nu _{2k}\right] }\,,
\m{contract}
\end{equation}%
we obtain for (\ref{NL1-lock}):
\be
{\bm \omega}^{\rho\lambda|\mu\nu}_{l1} = {\bm \omega}^{\rho\lambda|\mu\nu}_{h1} \l[\sum^{m}_{p=1}p\alf_p(-\ell_{eff}^{-2})^{p-1}\frac{(D-3)!}{(D-2p-1)!} \r]\,.
\m{NL1-AdS}
\ee
One easily finds (in this relation, see \cite{Rodrigo_2017}) that the expression in square brackets is defined by the differentiation of (\ref{sum}):
\be
 V'(\ell_{eff}^{-2}) = \l.\l(\di_x V(x)\r)\r|_{x =\ell_{eff}^{-2} } = \sum^{m}_{p=1}p\alf_p(-\ell_{eff}^{-2})^{p-1}\frac{(D-3)!}{(D-2p-1)!}\,.
\m{NL1-AdS+}
\ee
The expression (\ref{NL1-AdS}) shows that all the quantities (\ref{Linear-Genl}) - (\ref{Super-l}), if they are constructed for the AdS background that is the vacuum one, are proportional to the factor (\ref{NL1-AdS+}). The role of the coefficient (\ref{NL1-AdS+}) is discussed in detail at the end of the present section.

Now, making use of (\ref{NL1-AdS}) with (\ref{NL1-H}), we get the following important expressions.
First, the linear operator (\ref{Linear-Genl}) under the condition (\ref{R_AdS}) acquires the form:
\bea
&& {\cal G}^{\mu\nu}_L = \frac{\sqrt{-\bar g}}{2} V'(\ell_{eff}^{-2})\l[\bar \nabla_{\rho}{}^\mu\varkappa^{\nu\rho} + \bar \nabla_{\rho}{}^\nu\varkappa^{\mu\rho} - \bar \nabla_{\rho}{}^\rho\varkappa^{\mu\nu} - \bar g^{\mu\nu}\bar\nabla_{\rho\lam}\varkappa^{\rho\lam}\r.\nonumber\\ &&\l. + \bar g^{\mu\nu}\nabla_{\rho}{}^\rho\varkappa - \nabla^{\mu\nu}\varkappa + \bar g^{\mu\nu} \frac{2\Lambda_{eff}}{D-2}\varkappa -  \frac{4\Lambda_{eff}}{D-2}\varkappa^{\mu\nu}\r] \,.  \label{Linear_final}
\eea
Second, the same expression (\ref{Linear_final}) divided by $\k$ is related to the energy-momentum ${\bm \tau}^{\mu\nu}_{\ell}$ in (\ref{theta}) that is conserved:
\be
\bar\nabla_\nu{\bm \tau}^{\mu\nu}_{\ell} = 0.
\m{tau_conserve}
\ee
For (\ref{NL1-AdS}) with (\ref{NL1-H}) the conserved current (\ref{current-l}) is calculated with making the use of ${\bm \tau}^{\mu\nu}_{\ell}$ and with $z$-term (\ref{ZDmu-l}) that can be easily found  as well.
Third, for (\ref{NL1-AdS}) with (\ref{NL1-H}) the superpotential (\ref{Super-l}) acquires the concrete form:
\be
{\cal I}^{\mu\nu}_{\ell} = \frac{\sqrt{-\bar g}}{\k} V'(\ell_{eff}^{-2})
 \l[\xi_\rho \bar\nabla^{[\mu}\varkappa^{\nu]\rho} - \xi^{[\mu} \bar\nabla_\rho \varkappa^{\nu]\rho}  + \xi^{[\mu} \bar\nabla^{\nu]}\varkappa
+ \varkappa^{\rho[\mu}\bar\nabla^{\nu]}\xi_\rho + \half \varkappa \bar\nabla^{[\mu} \xi^{\nu]}\r]\, .
 \m{DTsuperpotential}
 \ee

To check the above results we consider static black holes in the Lovelock gravity. To have concrete formulae for the solution we use the presentation given in \cite{Kofinas_Olea_2007}. As usual, a static black hole solution can be presented by the Schwarzschild-like metric:
 \be
 d s^2 = -fdt^2 + \frac{1}{f}dr^2 +
 r^2\sum_{a,b}^{D-2}q_{ab}dx^adx^b\, ,
 \m{AdS}
 \ee
 where
 $f$ satisfies the equation
 \be
 \sum^{m}_{p=0}\frac{\alf_p}{(D-2p-1)!}\l(\frac{1-f}{r^2} \r)^p = \frac{\mu}{(D-3)!\,r^{D-1}}
 \m{EQ_rr}
 \ee
  which is a result of integration of the $rr$-component of the Lovelock vacuum equations with the constant of integration $\mu$. For the black hole solution one has to find the event horizon $r_+$ that is the largest solution of the equation $f(r_+) = 0$. We assume that such a solution exists, see \cite{Kofinas_Olea_2007}. Besides, in \cite{Kofinas_Olea_2007} it is shown that the asymptotic behaviour of $f$ at $r\goto\infty$ is
 \be
 f(r)\sim 1 +\frac{r^2}{\ell^2_{eff}} - \frac{1}{V'(\ell_{eff}^{-2})}\frac{\mu}{r^{D-3}}\,.
 \m{f_goto}
 \ee
Comparing it with (\ref{fBar}), one gets
 \be
  \Delta f = f(r) - \bar f(r) \sim - \frac{1}{V'(\ell_{eff}^{-2})}\frac{\mu}{r^{D-3}}\,.
 \m{f_delta}
 \ee
 As a result, for the perturbations (\ref{h_a}) one has
 \be
 \varkappa_{00} \sim -\Delta f,\qquad \varkappa_{11} \sim -\Delta f/\bar f^2
 \m{h_00_11}
 \ee
in the appropriate order of approximation.

To calculate the mass for the black hole solution (\ref{AdS}) with the AdS asymptotic (\ref{AdS_back}) one has to use the Killing vector $\bar\xi^\alf = \{-1,0,0,0 \}$ and $01$-component of the superpotential (\ref{DTsuperpotential}) with the appropriate order of approximation for the perturbations (\ref{h_00_11}). Thus with making the use of (\ref{ChristAdS}) we obtain
\be
{\cal I}^{01}_{\ell} \sim -\frac{\sqrt{-\bar g}}{\k} V'(\ell_{eff}^{-2}) \frac{D-2}{2r} \Delta f\,.
\m{SUP_01}
\ee
Then, substituting (\ref{f_delta}) and taking into account $\sqrt{-\bar g} = r^{D-2}\sqrt{\det q_{ij}}$, one obtains for the mass
\be
M = \lim_{r \goto \infty} \oint d x^{D-2} {\cal I}^{01}_{\ell} =  \frac{D-2}{2\k }\mu\oint d x^{D-2}\sqrt{\det q_{ij}}= \frac{D-2}{4 G_D}\mu  \,.
\m{M}
\ee
It is the standard  accepted result obtained using
various approaches in various $D$-dimensional modifications of GR: for example, in \cite{DT1,DT2} in quadratic in curvature gravities; in \cite{AFrancavigliaR} in Chern-Simons and Lovelock gravity; in \cite{Okuyama} in higher curvature
gravitational theories; in \cite{Paddila} with using the Hamiltonian approach and constructing related surface terms in the EGB gravity; in \cite{DerKatzOgushi} in the EGB gravity using canonical approach developed in \cite{KBL}, {\em etc}, and numerous references therein.
One has to pay especial attention to the result (\ref{M}) obtained in  \cite{Kofinas_Olea_2007} with using the technique of counterterms, see, for example, \cite{Olea_2005,Olea_2007,Miskovic_Olea_2007} and references therein.

Finally, let us discuss the role of the coefficient (\ref{NL1-AdS+}) in the above formulae. The case, when $V'(\ell_{eff}^{-2}) \neq 0$ is the condition when the AdS background (AdS vacuum) is not degenerated. In this case one can use all the above formulae without obstacles. The case of a degenerate AdS vacuum,
\be
V'(\ell_{eff}^{-2}) = 0\,,
\m{VVV}
\ee
is analyzed in detail in \cite{Rodrigo_2017}, besides, the case of degenerate vacuum with multiplicity, like $V''(\ell_{eff}^{-2}) = 0, ~V'''(\ell_{eff}^{-2}) = 0$, etc., is analyzed as well.

In the case of a degenerate AdS background (\ref{VVV}), the field-theoretical current (\ref{current-l}) and superpotential (\ref{Super-l}) become zero for arbitrary perturbations. This could mean that they cannot describe a physical situation, or a degenerate background is not physical one. The linear operator (\ref{Linear_final}) becomes zero as well. Consequently, the total energy-momentum for perturbations at the right hand side of (\ref{EqsVacuum}) disappears also. This could be interpreted as a case where the matter energy-momentum is compensated by the pure gravitational energy-momentum. Such a situation looks strange. However, for example, this is an acceptable conclusion when the total energy of the closed Friedman world is equalized to zero where the matter energy is compensated by the gravitational energy \cite{Misner_Thorn_Wheeler_1973}.  In the case of absence of matter in (\ref{EqsVacuum}), the energy-momentum of the metric perturbations must to be equal to zero, and can be interpreted as absence of gravitons! Such a situation is discussed in detail in the EGB gravity in the paper \cite{Petrov2009}, where the condition analogous to (\ref{VVV}) has been analyzed without definite conclusions in the framework of various approaches, not only in the field-theoretical one.

It seems that in the case of calculating the charge (\ref{M}) of the Lovelock black hole the degeneracy condition (\ref{VVV}) does not lead to a problem because the degeneracy coefficient is cancelled in (\ref{SUP_01}). However, it is not so. Indeed, in \cite{Rodrigo_2017} the authors show that for the degenerate AdS background the approximation (\ref{f_goto}) does not hold: it has a slower fall-off and other coefficients. Thus,  in the case of the degeneracy condition (\ref{VVV}), the field-theoretical approach does not have a well-defined charge for the Lovelock black hole. In this regard, it is important to turn to the study in \cite{Rodrigo_2017}, where the authors develop the definition of the conformal mass  within a given AdS branch of the Lovelock gravity. The notion of conformal mass has been elaborated for the Einstein gravity in \cite{Ashtekar_Magnon_1984,Ashtekar_Das_2000}, where conserved quantities in asymptotically AdS spaces are encoded in the Weyl tensor. This is quite natural, as it is a tensor that captures the conformal properties of the spacetime. So, in \cite{Rodrigo_2017} conserved charges have been obtained with making the use of the counterterm technique \cite{Olea_2005,Olea_2007,Miskovic_Olea_2007}. The charges turn out to be proportional to the electric part of the Weyl tensor with a factor of proportionality as the degeneracy condition. Thus, the conformal mass is well-defined in the case of absence of the degeneracy, and has the obstruction in the degenerate case, the same as in our formulae above.
The authors of \cite{Rodrigo_2017} conclude that this fact reflects an obstruction to the linearization of the theory. As a consequence, in the degenerate case, the information on the black hole mass has to be contained in non-linear terms of the boundary energy-momentum tensor.

\section{Discussion}
 \m{Discussion}
\setcounter{equation}{0}

Let us summarize, remark on novelty and discuss the importance of our central results.

First, for the Lovelock gravity we have obtained currents in a concrete form (\ref{current-l}) for perturbations on {\em arbitrary curved} backgrounds (not only vacuum ones) of   all the permissible types satisfying (\ref{bbb}). Their structure is determined by the total energy-momentum, ${\bm \tau}^{\mu\nu}_l$, and z-term, ${\bm z}^{\mu}_l$. The last is  given in (\ref{ZDmu-l}) and disappears if displacement vectors are Killing ones, $\xi^\alf = \bar\xi^\alf$, of the background. The total energy-momentum is analyzed in detail. For complicated backgrounds ${\bm \tau}^{\mu\nu}_l$ has a general form (\ref{EM1+}), or (\ref{EM1++A}), whereas for a vacuum case it is (\ref{EM1++B}). Besides, in the case of a given solution one can use the expression (\ref{theta}).

Second, for the Lovelock gravity, superpotentials in a concrete form (\ref{Super-l}) have been obtained for perturbations on {\em arbitrary curved} backgrounds (not only vacuum ones) of all the permissible types (\ref{bbb}). We remark that the form of superpotentials (\ref{Super-l}) is unique and universal one (\ref{(+16+A)}) for all the cases considered here.

We note that a consideration of non-vacuum backgrounds is very important because many physically interesting solutions are just non-vacuum ones, for example, the Friedmann solution. Thus, in the framework of the Einstein gravity, in the series of the works \cite{PK1,PK2,PK3,PK4}, we develop a derivation of cosmological perturbations on the basis of the Lagrangian based field-theoretical method, see also chaper 5 of the book \cite{Petrov+_2017}.

Third, our results are valid in the case of arbitrary displacement vectors. This is important because conserved quantities can be constructed, for example, for proper vectors of observers which can be different from Killing vectors; or for {\em conformal Killing vectors} (these are not Killing vectors either), see  \cite{PK} and references there in; or for the {\em Kodama vectors} (these are not Killing vectors), see \cite{Kodama}, etc.

Fourth, we stress that our results are valid for arbitrary definitions of metric perturbations $h^a$ from the set (\ref{Phi-Dec}), see (\ref{B40}) as well. It is important for the reasons given in the last paragraph of section \ref{PerturbedDTheory}.

Fifth, calculation of the mass of static black holes with the AdS asymptotic in the Lovelock gravity plays a role of a test for the developed formalism. Future applications are discussed below.

Let us look at the field-theoretical presentation at the level of the equations for perturbations on given backgrounds. Such gravitational equations are derived in the form where the linearized metric part ${\cal G}^L_{\mu\nu}$ and linearized matter part ${\cal F}^L_{\mu\nu}$ are placed on the left hand side, whereas all the other terms are placed on the right hand side and treated as an effective (total) energy-momentum ${\bm t}^{tot}_{\mu\nu}$. All approaches, where such equations are analyzed, we classify as a field-theoretical method in general. Among various approaches there is the {\em Lagrangain based} one where the equations ${\cal G}^L_{\mu\nu} + {\cal F}^L_{\mu\nu} = \k{\bm t}^{tot}_{\mu\nu}$ are obtained by varying a related Lagrangian, in contrast in other approaches where these equations are obtained 'by hand' after a direct linearization. In the {\em Lagrangain based} formulation perturbations play a role of dynamical variables, we call a related Lagrangian as the dynamic one, and here it is defined in (\ref{lag}).

In the present paper, we apply the {\em Lagrangian based} method only. Its advantages follow from the fact that conserved quantities are obtained by exploiting symmetries of the dynamic Lagrangian using the Noether theorem, like in the usual covariant field theory on a fixed background. To get a feel for these advantages one has to compare the Lagrangian based method with another one where a dynamic Lagrangian or its analogies are not used. Among such approaches possibly the best known, fruitful and popular  is the field-theoretical approach by Deser and Tekin and their coauthors (see, for example, \cite{DT1,DT2,DT3,DT4,DT5}). This approach applies the Abbott
and Deser procedure developed for 4D GR \cite{AbbottDeser82} in metric of higher curvature gravity theories in $D$ dimensions, and frequently is called as the ADT approach. Its intensive development, and many applications have many important results; for any generic f(Riemann) theory with constructed ADT charges, including the Lovelock
theory, see \cite{DT6}, for a wide review see chapter 9 of the book \cite{Petrov+_2017} and the recent review \cite{DT7}. Below we provide aforementioned comparison.

  First, by the ADT method, one provides a direct linearization  without varying Lagrangians. On the other hand, the Lagrangian based method permits to derive linear parts of the gravitational equations making use of the definitions in (\ref{GL-q}) and (\ref{PhiL-q}) that plays a role of universal algorithm.

 Second, to the best of our knowledge, perturbations in the framework of the ADT approach are considered on vacuum (as a rule, AdS vacuum) backgrounds only. At the same time, the Lagrangian based method permits construction of conserved quantities on arbitrary curved backgrounds of all the permissible types satisfying (\ref{bbb}).

Third, for vacuum backgrounds the linearization of matter part disappears, ${\cal F}^L_{\mu\nu}= 0$. Thus, the gravitational equations in the field-theoretical presentation become ${\cal G}^L_{\mu\nu} = \k{\bm t}^{tot}_{\mu\nu}$. Moreover, in the vacuum case, $\bar\nabla^\nu{\cal G}^L_{\mu\nu}\equiv 0$. Then, for the ADT approach this identity is contracted with a Killing vector $\bar \xi^\mu$ that gives the identity $\bar\nabla^\nu\l({\cal G}^L_{\mu\nu}\bar \xi^\mu\r)\equiv 0$. This signals that a related superpotential has to exist and, indeed, it is constructed in the framework of the ADT approach. However, the ADT procedure does not work if non-Killing vectors are used. In this case, one needs z-term drastically, but it is not determined by the ADT procedure at all.  On the other hand, the Lagrangian based method permits construction of conserved quantities even for non-Killing displacement vectors.

Fourth, in the case of non-vacuum backgrounds one has $\bar\nabla^\nu{\cal G}^L_{\mu\nu}\neq 0$ and $\bar\nabla^\nu\l({\cal G}^L_{\mu\nu} + {\cal F}^L_{\mu\nu}\r)\neq 0$, which is  due to an interaction of perturbations with complicated backgrounds. As a result, the ADT procedure fails in constructing conserved quantities on non-vacuum backgrounds. On the other hand, the Lagrangian based method permits construction of conserved currents and suprpotentials even on non-vacuum backgrounds. As a result, the interacting terms are defined and incorporated into the energy-momentum in the explicit form, see (\ref{EM1+}) and (\ref{EM1++A}).

 Fifth,  in the ADT formalism, one actually directly obtains conserved currents because of the contraction of the field equations with a Killing vector field. Therefore, in the ADT approach, one really does not construct a conserved energy-momentum rank-2 tensor. In the Lagrangian based formalism, one has a little more: one presents a total and detailed structure of a conserved current including the structure of the energy-momentum tensor. It is quite natural because from the start the Lagrangian based method has been suggested for describing local characteristics for perturbations propagating on a given background. Of course, on the vacuum background and after choosing Killing vectors we get the ADT form.

Sixth, both the Lagrangian based expressions (in the case of vacuum backgrounds and Killing vectors) and the ADT expressions for the charges are gauge invariant up to boundary terms only,  which can vanish or not for different boundary conditions, see chapters 3 and 9 in the book \cite{Petrov+_2017}, respectively. Recently, this situation is remedied, and explicitly gauge-invariant conserved quantities are obtained in GR \cite{AT1} and in a generic theory \cite{AT2}.

Now we return to a discussion of the results of the present paper themselves. Our technique essentially is based on using the auxiliary Lagrangians $\lag_1$ of the type (\ref{Lag-1}) that can be related the Lovelock theory only, see the first paragraph of section \ref{ConservedCurrents}. Of course, for arbitrary higher curvature gravity theories in $D$ dimensions, like quadratic gravity, see \cite{DT1,DT2}, or the $f(R)$ theory, see \cite{Sotiriou}, etc., application of the Lagrangian based field-theoretical method is possible. However the remarkable properties of the expressions (\ref{current-l}) and (\ref{Super-l}) is destroyed. Indeed, for more complicated theories the right hand side of (\ref{Lag-1}) can depend on even fourth derivatives of the background metric. Then one needs significantly more complicated Klien-Noether systems of identities than (\ref{(+9+1)}) - (\ref{(+9+4)}), and, in this case, just the quantity, like (\ref{NL1}), is not enough to describe the system.

Finally, let us discuss possibility of choosing an acceptable gravitational variable $h^a$ for definite goals among various metric perturbations defined in (\ref{g-Dec}).
The difference in (\ref{B40}) is not important for calculating conserved quantities for static solutions. Therefore we use the definition (\ref{h_a}) as a more simple one for calculations in sections \ref{Lovelock} and \ref{S-AdS-mass}. It was shown in 4D GR that for a radiating isolated system a different choice of the metric perturbations gives, see \cite{PK}, different results for the Bondi-Sachs momentum \cite{BMS}. Thus, it would be useful to apply the results of the present paper with different choices of perturbations to analyze radiating solutions in the Lovelock theory.

Of course, the listed above advantages of the presented approach can be (and must be) applied to study solutions in the Lovelock gravity, which cannot be examined with using other methods. In Introduction, we have remarked that the Lovelock theory is quite popular and important now. It is the most natural generalization when one comes to higher dimensions. Moreover, there are arguments that only a so-called {\em pure} Lovelock gravity leads to acceptable  equations in higher dimensions, see, for example, \cite{Dadhich+_2012,Dadhich+_2013,Dadhich_2016}. The pure Lovelock gravity is presented by a one polynomial only from the sum of all the polynomials in the total Lovelock Lagrangian.

Therefore, because there is a definite interest in the pure Lovelock gravity, in future we plan to apply our results to study solutions of this theory obtained in \cite{Dadhich+_2013}. The first one presents a collapsing inhomogeneous dust. By our opinion, it is quite interesting and important to examine stability/instabilty of this solution. In another word, one needs to study perturbations, their characteristics and evolution on a background of this solution itself. Because this background is non-vacuum (with matter) our approach looks as a qiute appropriate one to be applied. On the other hand, approaches constructed for maximally symmetric backgrounds, like the ADT one, cannot be used in this case. The second solution in \cite{Dadhich+_2013} presents the Vaidya-type collapsing/radiating model with light-like matter (null dust). To understand this model deeper it is important to study densities of conserved quantities measured by a system of observers. Our method is quite appropriate for such a study. Indeed, first, initially the field-theoretical method has been elaborated for studying local characteristics; second, in constructing aforementioned local densities proper vectors of observers (which are not Killing vectors in general) have to be used. It is permissible for our method, although other approaches, like the ADT one, use Killing vectors only.

 Another useful application that could be carried out is as follows. It is noted in \cite{Dadhich+_2012,Dadhich+_2013,Dadhich_2016} that a pure Lovelock gravity in even dimensions has properties which are very close to those in 4-dimensional Einstein theory. Therefore, it could be interesting to represent, for example, 6 dimensional the pure Lovelock theory to the field-theoretical form and compare with the field-theoretical reformulation of 4 dimensional Einstein theory that already has been developed in detail, see Chapter 2 in the book \cite{Petrov+_2017} and references therein.

\bigskip

\noindent {\bf Acknowledgments}
The author is very grateful to Rodrigo Olea for discussions and fruitful recommendations; he thanks Nathalie Deruelle for good suggestions to improve presentation; he thanks Bayram Tekin for useful discussions, recommendations and explanation of his with coauthors works; he thanks Naresh Dadhich for discussion of possible applications of the presented formalism; and he thanks Deepak Baskaran for improving English. The author acknowledges also the support from the Program of development of M.V. Lomonosov Moscow State University (Leading Scientific School 'Physics of stars, relativistic objects and galaxies').

\ed
\begin{thebibliography}{999}


\bibitem{Lovelock} Lovelock D 1971 The Einstein tensor and its
generalizations {\em J. Math. Phys.} {\bf 12} 498

\bibitem{Troncoso+_2000} Crisostomo J, Troncoso R and Zanelli J 2000 Black Hole Scan {\em Phys. Rev. D} {\bf 62} 084013  arXiv:hep-th/0003271

\bibitem{Cai_Ohta_2006} Cai R-G, Ohta N 2006 Black Holes in Pure Lovelock Gravities {\em Phys. Rev. D} {\bf 74} 064001 arXiv:hep-th/0604088

\bibitem{Taves_2014} Taves T 2014 Black Hole Formation in Lovelock Gravity {\em Preprint (PhD thesis)} arXiv:1408.2241 [gr-qc]

\bibitem{Aros_Estrada_2019} Aros R and Estrada M 2019 Regular black holes and its thermodynamics in Lovelock gravity arXiv:1901.08724 [gr-qc]

\bibitem{Pavluchenko_Toporensky_2014} Pavluchenko S and Toporensky A 2014 Note on properties of exact cosmological solutions in Lovelock gravity
{\em Gravit. Cosmol.} {\bf 20} 127 arXiv:1212.1386 [gr-qc]


\bibitem{Petrov2009} Petrov A N 2009 Three types of superpotentials
for perturbations in the Einstein-Gauss-Bonnet gravity {\em Class.
Quantum Grav.} {\bf 26} 135010; Corrigendum: {\em Class.
Quantum Grav.} {\bf 27} (2010) 069801  arXiv:0905.3622 [gr-qc]

\bibitem{Petrov2009a} Petrov A N 2010 On creating mass/matter by extra
dimensions in  the Einstein-Gauss-Bonnet gravity {\em Grav. Cosmol.}
{\bf 16} 3 arXiv:0911.5419 [gr-qc]

\bibitem{Petrov2011} Petrov A N 2011 Noether and Belinfante
corrected types of currents for perturbations in the
Einstein-Gauss-Bonnet gravity  {\em Class. Quantum Grav.} {\bf 28}  215021 arXiv:1102.5636 [gr-qc]

\bibitem {KBL} Katz J, Bi\v c\'ak J and  Lynden-Bell D 1997
Relativistic conservation laws and integral constraints for large
cosmological perturbations {\em Phys. Rev. D} {\bf 55} 5957 arXiv:gr-qc/0504041

\bibitem{PK} Petrov A N and  Katz J 2002
Conserved currents, superpotentials and cosmological perturbations
{\it Proc. R. Soc. A, London} {\bf 458} 319   arXiv:gr-qc/9911025

\bibitem{Weinberg-book} Weinberg S 1972
{\em Gravitation and Cosmology} (Wiley, New York)

\bibitem{[11]}  Deser S 1970 Self-interaction and gauge invariance
{\em Gen. Relat. Grav.} {\bf 1} 9 arXiv:gr-qc/0411023

\bibitem{Deser87} Deser S 1987 Gravity from self-interaction
in a curved background {\em Class. Quantum Grav.} {\bf 4} 99

\bibitem{GPP} Grishchuk L P, Petrov A N  and Popova A D 1984
Exact theory of the (Ein\-stein) gravitational field in an arbitrary
background space-time {\em Commun. Math. Phys.} {\bf 94} 379

\bibitem{[15]} Popova A D and   Petrov A N 1988
The  dynamic  theories  on  a fixed  background in gravitation {\em
Int. J. Mod. Phys. A} {\bf 3} 2651

\bibitem{GP87} Grishchuk L P and Petrov A N 1987 The Hamiltonian description
of the gravitational field and gauge symmertries {\em Sov. Phys.:
JETP} {\bf 65} 5 [1986 {\em Zh. Eksp. Teor. Fiz.} {\bf 92}, 9]

\bibitem{CQG_DT}  Petrov A N 2005 A note on the Deser-Tekin charges {\em Class. Quantum Grav.} {\bf 22} L83 arXiv:gr-qc/0504058

\bibitem{Petrov2008} Petrov A N 2008 Nonlinear perturbations and conservation laws on
curved backgrounds in GR and other metric theories, 2-nd chapter in
the book: {\em Classical and Quantum Gravity Research} Eds:
Christiansen M N and Rasmussen T K (Nova Science Publishers, N.Y.)
79 arXiv:0705.0019 [gr-qc]

\bibitem{Petrov+_2017}	Petrov A N, Kopeikin S M, Lompay R R and Tekin B 2017
{\em Metric Theories of Gravity: Perturbations and Conservation Laws} (de Gruyter: Germany)

\bibitem{Szabados_2009} Szabados L B 2009 Quasi-local energy-momentum and angular momentum in general relativity
{\em Living Reviews in Relativity} {\bf 12}

\bibitem{Petrov_Lompay_2013} Petrov A N and Lompay R R 2013 Covariantized Noether identities and conservation laws
for perturbations in metric theories of gravity {\em Gen. Relat. Grav.} {\bf 45} 545 arXiv:1211.3268 [gr-qc]




\bibitem{B-Deser} Boulware D C and  Deser S 1975 Classical general
relativity derived from quantum gravty {\it Ann. Phys.} {\bf 89},
93

\bibitem{BMS} Bondi H,  Metzner A W K and Van der Berg M J C 1962
Gravitational waves in general relativity. VII. Waves from
axi-symmetrical isolated systems  {\em Proc. R. Soc. A London} {\bf
269} 21

\bibitem{AbbottDeser82}
Abbott L F  and Deser S 1982 Stability of gravity with a
cosmological constant  {\it Nucl. Phys. B} {\bf 195} 76

\bibitem{Mitzk} Mitzkevich N V 1969  {\em Physical Fields in
General Theory of Relativity} (Nauka, Moscow), in Russian


\bibitem{DT1} Deser S and Tekin B 2002 Gravitational energy in
quadratic curvature gravities {\it Phys. Rev. Lett.} {\bf 89} 101101 arXiv:hep-th/0205318

\bibitem{DT2} Deser S and Tekin B 2003 Energy in generic
higher curvature gravity theories {\em Phys. Rev. D} {\bf 67} 084009 arXiv:hep-th/0212292

\bibitem{Sotiriou} Sotiriou T P and Faraoni V 2010 $f(R)$ theories of gravity
{\em Rev. Mod. Phys.} {\bf 82} 451 arXiv:0805.1726 [gr-qc]


\bibitem{Komar59} Komar A 1959
{\it Phys. Rev.} {\bf 113}, 934



\bibitem{Lompay_Petrov_2013a} Lompay R R and Petrov A N 2013 Covariant Differential Identities and Conservation Laws in Metric-Torsion Theories of Gravitation. I. General Consideration   {\em  J. Math. Phys.} {\bf 54} 062504  arXiv:1306.6887 [gr-qc]

\bibitem{Lompay_Petrov_2013b} Lompay R R and Petrov 2013 Covariant Differential Identities and Conservation Laws in Metric-Torsion Theories of Gravitation. II. Manifestly generally covariant theories  {\em J. Math. Phys.} {\bf 54}  102505  arXiv:1309.5620 [gr-qc]

\bibitem{Rund} Rund H 1966 Variational problems involving
combining tensor fields {\em Abhandl. Math. Sem. Univ. Hamburg} {\bf
29} 244

\bibitem{Olea_2005} Olea R 2005   Mass, Angular Momentum and Thermodynamics in Four-Dimensional Kerr-AdS Black Holes
{\em JHEP} {\bf 0506} 023 arXiv:hep-th/0504233


\bibitem{Giacomini+1} Canfora F, Giacomini A and Pavluchenko S A 2013 Dynamical compactification in Einstein-Gauss-Bonnet gravity from geometric frustration
{\em Phys. Rev. D} {\bf 88} 064044 arXiv:1308.1896 [gr-qc]

\bibitem{Giacomini+2} Chirkov D, Giacomini A and Toporensky A 2018 Dynamic compactification with stabilized extra dimensions in cubic Lovelock gravity
arXiv:1804.02193 [gr-qc]


\bibitem{Rodrigo_2017}   Arenas-Henriquez G, Miskovic O and Olea O 2017 Vacuum Degeneracy and Conformal Mass in Lovelock AdS Gravity {\em JHEP} {\bf 1711} arXiv:1710.08512 [hep-th]

\bibitem{Kofinas_Olea_2007} Kofinas G and Olea R 2007 Universal regularization prescription for Lovelock AdS gravity
{\em  JHEP} {\bf 0711} 069 arXiv:0708.0782 [hep-th]







\bibitem{AFrancavigliaR} Allemandi G, Francaviglia M and Raiteri M
2003 Charges and energy in Chern-Simons theories and Lovelock
gravity {\em Class. Quantum Grav}. {\bf 20} 5103 arXiv:gr-qc/0308019

\bibitem{Okuyama} Okuyama N and Koga J-I 2005 Asymptotically anti-de
Sitter spacetimes and conserved quantities in higher curvature
gravitational theories {\em Phys. Rev. D} {\bf 71}  084009  arXiv:hep-th/0501044

\bibitem{Paddila} Paddila A 2003 Surface terms and
Gauss-Bonnet Hamiltonian {\it Class. Quantum Grav.} {\bf 20} 3129 arXiv:gr-qc/0303082

\bibitem{DerKatzOgushi} Deruelle N, Katz J and Ogushi S 2004 Conserved
charges in Einstein-Gauss-Bonnet theory {\it Class. Quantum Grav.}
{\bf 21} 1971 arXiv:gr-qc/0310098




\bibitem{Olea_2007} Olea R 2007 Regularization of odd-dimensional AdS gravity: Kounterterms
{\em JHEP} {\bf 0704} 073 arXiv:hep-th/0610230

\bibitem{Miskovic_Olea_2007} Miskovic O and Olea R 2007 Counterterms in Dimensionally Continued AdS Gravity
{\em JHEP} {\bf 0710} 028 arXiv:0706.4460 [hep-th]




\bibitem{Misner_Thorn_Wheeler_1973} Misner C W, Thorne K S and Wheeler J A 1973 {\em Gravitation} (Freeman W H and Company, San Francisco)


\bibitem{Ashtekar_Magnon_1984} Ashtekar A  and  Magnon A 1984 Asymptotically anti-de Sitter space-times {\em Class. Quant. Grav.}
{\bf 1}  L39

\bibitem{Ashtekar_Das_2000} Ashtekar A and Das S 2000 Asymptotically Anti-de Sitter space-times: Conserved quantities
{\em Class. Quant. Grav.} {\bf 17} L17  arXiv:hep-th/9911230



\bibitem{PK1}	Kopeikin S M and Petrov A N 2013 Post-Newtonian Celestial Dynamics in Cosmology: Field Equations {\em Phys. Rev. D}  {\bf 87} 044029 arXiv:1301.5706 [gr-qc]

\bibitem{PK2} Kopeikin S M and Petrov A N 2014 Dynamic Field Theory and Equations of Motion in Cosmology {\em Ann. Phys.} {\bf 350} 379  arXiv:1407.3846 [gr-qc]

\bibitem{PK3}	Petrov A N and Kopeikin S M  2014 {\em Post-Newtonian approximations in cosmology}; the chapter in the book: {\em Frontiers in Relativistic Celestial Mechanics, Volume 1: Theory, Series: De Gruyter studies in mathematical physics, v. {\bf 21}}, Ed.: Kopeikin S M (De Gruyter, Germany) 295--392

\bibitem{PK4}	Kopeikin S M and Petrov A N 2015 {\em Equations of Motion in an Expanding Universe}; the chapter in the book: {\em Equations of Motion in Relativistic Gravity,  Series: Fundamental Theories of Physics, v. {\bf 179}}, Eds.: Puetzfeld D, Lämmerzahl C and Schutz B (Springer International Publishing) 689--757

\bibitem{Kodama} Kodama H 1980 Conserved energy flux for the spherically symmetric system and the backreaction problem in the black hole evaporation {\em Prog. Theor. Phys.}
{\bf 63} 1217


\bibitem{DT3} Deser S, Kanik I and Tekin B 2005 Conserved charges in higher D Kerr-AdS spacetimes {\em Class.
Quantum Grav.} {\bf 22} 3383   arXiv:gr-qc/0506057

\bibitem{DT4} Deser S and Tekin B 2007 New energy definition for higher curvature gravities {\em Phys. Rev. D}
{\bf 75} 084032 arXiv:gr-qc/0701140

\bibitem{DT5} Deser S, Liu H, Lu H, Pope C N, Sisman T C and Tekin B 2011 Critical points of D-dimensional
extended gravities {\em Phys. Rev. D} {\bf 83} 061502  arXiv:1101.4009 [hep-th]

\bibitem{DT6} Senturk C, Sisman T C and Tekin B 2012 Energy and angular momentum in generic f(Riemann) theories {\em Phys. Rev. D} {\bf 86} 124030 arXiv:1209.2056 [hep-th]

\bibitem{DT7} Adami H, Setare M R, Sisman T C and Tekin B 2017 Conserved charges in extended theories of gravity arXiv:1710.07252  [hep-th]

\bibitem{AT1} Altas E and Tekin B 2019 Conserved charges in AdS: A new formula {\em Phys. Rev. D} {\bf 99} 044026 arXiv:1811.00370 [hep-th]

\bibitem{AT2} Altas E and Tekin B 2019 New approach to conserved charges of generic gravity in AdS {\em Phys. Rev. D} {\bf 99}  044016 arXiv:1811.11525 [hep-th]

\bibitem{Dadhich+_2012} Dadhich N, Ghosh S G  and Jhingan S 2012 The Lovelock gravity in the critical spacetime dimension {\em Phys. Lett. B} {\bf 711} 196 arXiv:1202.4575 [gr-qc]

\bibitem{Dadhich+_2013} Dadhich N, Ghosh S G  and Jhingan S 2013 Gravitational collapse in pure Lovelock gravity in higher dimensions
 {\em Phys Rev. D} {\bf 88} 084024  arXiv:1308.4312 [gr-qc]

\bibitem{Dadhich_2016} Dadhich N 2016 A discerning gravitational property for gravitational equation in higher dimensions
{\em Euro. Phys. J. C} {\bf 76} 104 arXiv:1506.08764 [gr-qc]














\end{thebibliography}
